# Multi-parameter constraints on empirical infrasound period–yield relations for bolides and implications for planetary defense

Elizabeth A. Silber[1,*], Josep M. Trigo-Rodríguez[2], Iyare Oseghae[3,†], Eloy Peña Asensio[4], Mark Boslough[5,6], Rodney Whitaker[5], Christoph Pilger[7], Philip Lubin[8], Vedant Sawal[1], Claus Hetzer[9], Randy Longenbaugh[1], Peter Jenniskens[10], Brin Bailey[8], Esther Mas Sanz[11], Patrick Hupe[7], Alexander N. Cohen[8], Thom R. Edwards[1], Sasha Egan[12], Reynold E. Silber[13], Summer Czarnowski[14,†], Miro Ronac Giannone[1]

[1]Sandia National Laboratories, 1515 Eubank Blvd NE, Albuquerque, NM, 87123, USA; [2]Institute of Space Sciences (ICE-CSIC/IEEC), Campus UAB, Barcelona, Catalonia, Spain; [3]University of Texas at San Antonio, 1 UTSA Circle, San Antonio, Texas, 78249, USA; [4]Dipartimento di Scienze e Tecnologie Aerospaziali, Politecnico di Milano (PoliMi), Milano, 20156, Lombardia, Italy; [5]Los Alamos National Laboratory, Los Alamos, NM, 87545, USA; [6]University of New Mexico, Albuquerque, NM, 87131, USA; [7]Federal Institute for Geosciences and Natural Resources (BGR), Hannover, Germany; [8]Physics Department, UC Santa Barbara, Santa Barbara, CA, 93106, USA; [9]National Center for Physical Acoustics, The University of Mississippi, University, MS 38677 [10]SETI Institute, CA, 94043, USA; [11]Institute of Environmental Engineering, Swiss Federal Institute of Technology (ETH), Zürich, Switzerland; [12]New Mexico Institute of Mining and Technology, Socorro, NM, USA; [13]Department of Biology, Chemistry and Environmental Sciences (BCES), Northern New Mexico College, 921 N. Paseo de Oñate, Española, NM, 87532, USA; [14]North Dakota State University, Fargo, ND, 58108, USA

*Corresponding author: esilbe@sandia.gov

†Was at Sandia when this work was underway.

**Abstract**

How effective are methods for estimating bolide energies from infrasound signal period-yield relationships? A single global period–energy relation can obscure significant variability introduced by parameters such as the atmospheric Doppler wind profile and the bolide's energy deposition profile as a function of altitude. Bolide speed, entry angle, burst altitude, and multi-episode fragmentation all may play a role in defining the detected period of the shockwave. By leveraging bolide light curve data from the Center for Near Earth Object Studies (CNEOS), we re-examined the period-energy relation as a function of these parameters. Through a bootstrap approach, we show that various event subsets can deviate from widely cited period–energy models and we identify which specific conditions most strongly reshape the period–energy scaling. The results define both the fidelity and reliability of period-energy relations when no additional data beyond the infrasound record is available and improve the outcome when supporting data from bolide trajectories and light curves are included. Ultimately, these findings expand the scope of earlier models, providing a nuanced and robust framework for infrasound-only yield estimation under a range of bolide scenarios.

## 1. Introduction

The need for simple, practical estimators of bolide energy using infrasound signals stretches back to methods originally developed for monitoring nuclear tests, most notably the Air Force Technical Applications Center (AFTAC) period–yield relations (Revelle, 1997). The AFTAC relations were established based on an analysis of the dominant period of nuclear explosion signals at source-to-station distances of 1300–8500 km (Glasstone and Dolan, 1977; Revelle, 1997).

The general form for period-based energy relations is:

$$\log(E) = A\log(P) + B \qquad (1)$$

where $E$ is energy in units of kilotons of TNT equivalent (1 kt = $4.184 \cdot 10^{12}$ J), $P$ is the dominant infrasound signal period (in seconds), typically measured at maximum amplitude, and $A$ and $B$ are the regression coefficients. Some relations employ an averaged signal period ($\bar{P}$), calculated by averaging the measured period across all stations that detected a particular high-energy event. Table 1 lists all empirical period-yield relations published to date, including those derived in this study.

**Table 1:** Compilation of empirical period–yield relation coefficients. The original Ens et al. (2012) relations were given in tons of TNT; here, we also present the corrected expressions in kilotons and address minor typographical errors that have appeared in the literature. $P*$ represents the signal period at each station ($P$) or the signal period averaged ($\bar{P}$) across all stations that detected a given event.

| A | B | P* | Units | Source | Notes |
|---|---|---|---|---|---|
| 3.30 (±0.14) | -1.89 (±0.10) | $P$ | kt | This study | -- |
| 3.71 (±0.42) | -2.07 (±0.21) | $\bar{P}$ | kt | This study | -- |
| 3.44 | -2.57 | $\bar{P}$ | kt | Whitaker (2023) | -- |
| 3.68 | -1.99 | $P$ | kt | Gi & Brown (2017) | -- |
| 3.84 | -2.21 | $\bar{P}$ | kt | Gi & Brown (2017) | -- |
| 3.75 | 0.50 | $P$ | t | Ens et al. (2012) | Original form |
| 3.75 | -2.50 | $P$ | kt | Ens et al. (2012) | Units corrected to kt |
| 3.28 | 0.71 | $\bar{P}$ | t | Ens et al. (2012) | Original form |
| 3.28 | -2.29 | $\bar{P}$ | kt | Ens et al. (2012) | Units corrected to kt |
| 3.26 | -2.40 | $P$ | kt | Whitaker and Mutschlecner (2006) | -- |
| 3.34 | -2.28 | $P$ | kt | ReVelle (1997) | $E < 200$ kt |
| 4.14 | -3.31 | $P$ | kt | ReVelle (1997) | $E > 80$ kt |

As applied to bolides, ReVelle's approach effectively reduced the total atmospheric energy by half, on the assumption that a significant fraction of the energy was radiated rather than contributing to the shock. This is reasonably in line with the energy partition estimates for meteoroid entries (Romig, 1965; Silber et al., 2018). Glasstone and Dolan (1977) state that this 50% partitioning only applies to bursts with altitudes <12 km and that at higher altitudes the fraction is lower. They also noted that the distribution between thermal radiation and blast changes more rapidly above 30 km, which could affect the energy-period relationship for events at very high altitudes. It is not well known, however, how this might apply to bolides. In Table 1, the first relations published by ReVelle (1997) are modified from their original form where the energy had been halved. In the original form, the coefficient $B$ is 0.3 smaller due to halving of the energy (i.e., the log(2) = 0.3). In this paper, all logs are assumed to be base 10.

In 2006, Whitaker and Mutschlecner (2006) used unclassified atmospheric nuclear explosion test data and derived a new relation, applicable to all energy ranges. This relation was recently updated by including an additional unclassified nuclear explosion dataset (Whitaker, 2023). These relations, because they are based on nuclear tests, assume a point source explosion, which differs from shock production by asteroids and large meteoroids (a cylindrical line source and/or a quasi-spherical source).

Studies by Ens et al. (2012) and Gi and Brown (2017), built on point-source explosions to produce empirical fits calibrated to bolide observations. Both of these studies utilized bolide events listed on the Center for Near-Earth Object Studies (CNEOS) fireball database, maintained by National Aeronautics and Space Administration's (NASA's) Jet Propulsion Laboratory (JPL). These bolides were detected by U.S. government (USG) sensors from space (Tagliaferri et al., 1994), providing details such as location, time and altitude of peak brightness. In some cases, the velocity vector is also available; this allows a tentative estimation of the heliocentric orbit (Peña-Asensio et al., 2022). Importantly, the CNEOS database includes total radiated energy (in joules) and impact energy (in kt) derived from flashes produced by bolides in the atmosphere and observed by USG space-based instruments in the silicon bandpass (Brown et al., 2002; Tagliaferri et al., 1994). The USG-derived energy estimates have been found to be robust (Borovička et al., 2015; Devillepoix et al., 2019; Wisniewski et al., 2024), thereby providing an important benchmark for validating and refining energy estimates derived from empirical infrasound-based relations. Currently, the light curves for the bolides that occurred prior to and including 2022 have been made available.

To derive period-based energy relations, Ens et al. (2012), compiled 63 CNEOS bolide events detected by 113 infrasound stations globally. They also used period measurements from multi-station detections of a single event to show that averaging can reduce scatter of data points (relative to the line defined by these relations). Their relations, updated here from their original form (Table 1) to reflect yields in kilotons and to correct minor typographical errors present in the published literature. The correction from ton to kilotons was done by reducing the original coefficient $B$ by 3 (i.e., log(1000) = 3). Remarkably, the relation with averaged periods was very similar to that by Revelle (1997).

Drawing on data from 78 CNEOS bolides detected by 179 stations, Gi and Brown (2017) established two additional relations, one that applies to all detections, and another that uses averaged periods for multi-station detections (Table 1). A portion of their dataset included bolides from Ens et al. (2012). Their fits from this larger dataset, however, are notably different, illustrating the need to rigorously evaluate the sensitivity of periods to energy and examine the robustness of period-based energy relations when applied to global bolide events. Although Gi and Brown (2017) examined a few bolides in greater detail by incorporating their light curves, neither they nor earlier studies integrated light curve analyses across a large number of events to directly connect energy-deposition modes with inferred energy estimates.

The fits from all these period-based energy relations are shown in **Figure 1**. As with any empirical regression, these relations differ in slope and intercept due to differing calibration datasets, some focusing on smaller bolide events, others on nuclear explosions, still others on partially idealized assumptions about the source function. Depending on the signal period length and the selected energy relation, the discrepancy in estimated yields can range from minor to over an order of magnitude. This outcome shows that both the choice of period–energy relation and whether data originate from single or multiple stations play a significant role in determining the final yield.

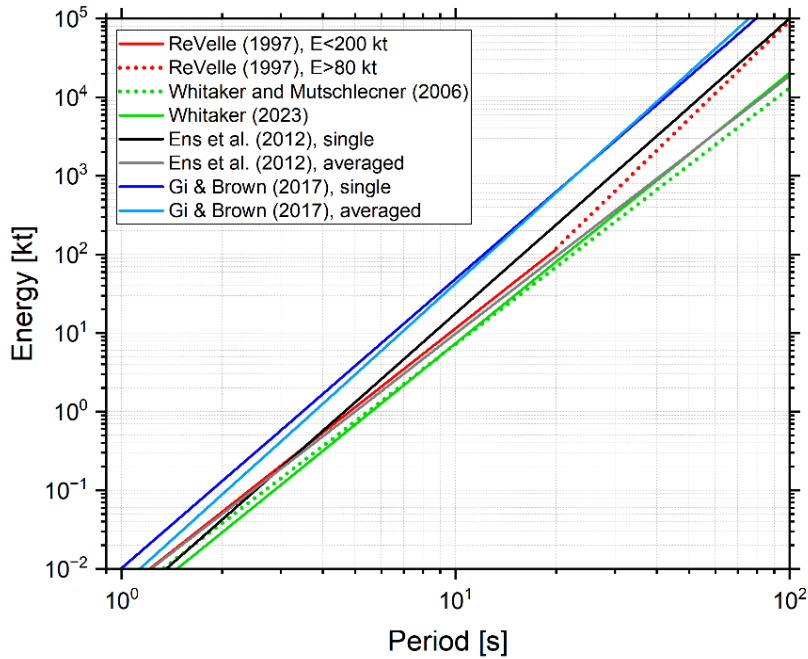

**Figure 1:** Log–log plot of energy versus signal period comparing published period–energy relations. Each line represents a different empirical fit: ReVelle (1997), Whitaker and Mutschlecner (2006), Whitaker (2023), and bolide-based relations from Ens et al. (2012) and Gi and Brown (2017), the latter two of which include both single-station and multi-station (averaged) data. Ens et al. (2012) and Gi and Brown (2017) calibrated the bolide energy relations against the USG sensor data, while the other relations stem from calibration against nuclear explosion data.

Many of the early infrasound-based yield estimations also incorporated amplitude relations, acknowledging that signal amplitude could, in principle, map to explosive source energy. However, amplitude-based fits typically demand substantial propagation corrections, particularly for range and atmospheric conditions, rendering them more sensitive to local winds, turbulence, and attenuation. By contrast, it can be argued that the signal period is more robust against these propagation effects, at least in the far field, and thus has served as the principal focus for most modern bolide energy studies. Accordingly, despite the inherent scatter in station-by-station period measurements, period-based relationships have remained the mainstay for rapid yield estimations.

The amplitude-based formulae are beyond the central scope of our paper. We include them for completeness and transparency in **Appendix A** (**Tables A1-3**). The tables consolidate the sometimes disparate approaches and correct for minor unit inconsistencies or typographical misprints that have appeared in past literature.

Over the last few decades, the empirical period–energy relations have formed the backbone of infrasound-based yield estimation, with the nuclear-explosion-derived relations (Revelle, 1997) historically serving as a simple, widely used benchmark. Because these period–energy fits require no detailed knowledge of event geometry or atmospheric profiles, they have remained attractive, despite recognized shortcomings such as station-dependent period variations (Ens et al., 2012). While subsequent studies (e.g., Ens et al., 2012; Gi & Brown, 2017) have expanded the event catalogs and offered more robust multi-station period averaging approaches, a universal 'global' regression still fails to adequately capture the complexity arising from more nuanced effects from atmospheric propagation and varied bolide deposition profiles. These complexities also reflect a broad range of meteorite properties, as meter-scale objects that generate noticeable infrasound signals frequently arrive at Earth already weakened by prior collisions (Beitz et al., 2016; Trigo-Rodriguez and Blum, 2009). Fragmentation can lead to increased ablation due to a higher surface/volume ratio.

Bolides can produce infrasound signals whose measured periods differ considerably from station to station, in part because each receiver can sample a distinct segment along the bolide path, each with potentially different effective burst altitude or local fragmentation (Silber, 2025; Silber et al., 2009). In principle, this would suggest the need to partition events not only by standard parameters like velocity or altitude, but also by more nuanced factors such as the number of fragmentation episodes in a single event, the density or dynamic strength, and the entry geometry (e.g., altitude, entry angle). Earlier works largely acknowledged this complexity (e.g., Ens et al., 2012; Gi and Brown, 2017), but typically did not incorporate it into a unified regression framework.

Amplitude-based relations for both artificial explosive sources (e.g., Mutschlecner and Whitaker, 2009; Pierce et al., 1971; Stevens et al., 2002) and bolides (Edwards et al., 2006; Ens et al., 2012) have also been developed, yet they suffer even more strongly from range and atmospheric effects (e.g., Stevens et al., 2002) and thus remain less reliable than period-based yield estimations. In light of these challenges, there is an obvious need to broaden the analysis of period–energy correlations using a more advanced framework that accounts for geometry (angle, altitude), physical parameters of bolides (mass, diameter), and light curve classification (single versus multiple 'bursts').

When a meteoroid or an asteroid enters Earth's atmosphere at velocities of tens of kilometers per second (11.2 – ~72 km/s), it travels far above the local speed of sound. As it descends into denser atmospheric layers, it becomes luminous and forms a shock wave (Bronshten, 1983; Ceplecha et al., 1998; Silber et al., 2018). Provided the object largely remains intact, this shock behaves much like a cylindrical line source, because the hypersonic bolide continuously deposits energy along its path. In such a regime, the shock front extends roughly perpendicular to the trajectory, and the pressure disturbances can be approximated as a cylindrical wave expanding outward from the flight path (Plooster, 1970; ReVelle, 1976). A useful parameter here is the blast radius ($R_0$) (Tsikulin, 1970). It defines the radial distance from the bolide trajectory within which the atmospheric pressure is significantly perturbed by the shock (Few, 1969). Closer to the meteoroid, the shock is strongly nonlinear (i.e., overpressures exceed ambient pressure by a large factor), whereas beyond the blast radius, the wave transitions into a more linear or weakly nonlinear regime (ReVelle, 1976).

As a first approximation, higher energy deposition yields a larger blast radius, shifting the shock signature toward lower frequencies (longer periods). However, in practice, the picture is more complicated. Atmospheric ablation can alter meteoroid flow regimes (Moreno-Ibáñez et al., 2018; Silber et al., 2018), and fragmentation events release energy in short segments or discrete 'bursts' (e.g., Trigo-Rodríguez et al., 2021), causing the shock wave to locally transition from a cylindrical (line-source) to a quasi-spherical (point-source) geometry (Silber and Brown, 2019). In a continuous flight with no major fragmentation, the shock predominantly resembles a cylindrical wave trailing behind the bolide. In contrast, an abrupt fragmentation (or airburst) produces a dominant near-spherical expansion, and multi-fragmentation events generate multiple overlapping quasi-spherical shock regions along the trajectory, complicating how the overall energy is distributed in time and space. In some instances, the observed light curve represents the sum of light curves generated by individual 'daughter' objects produced during a cascading fragmentation event. We discuss these complexities further in Section 4, in the context of our results.

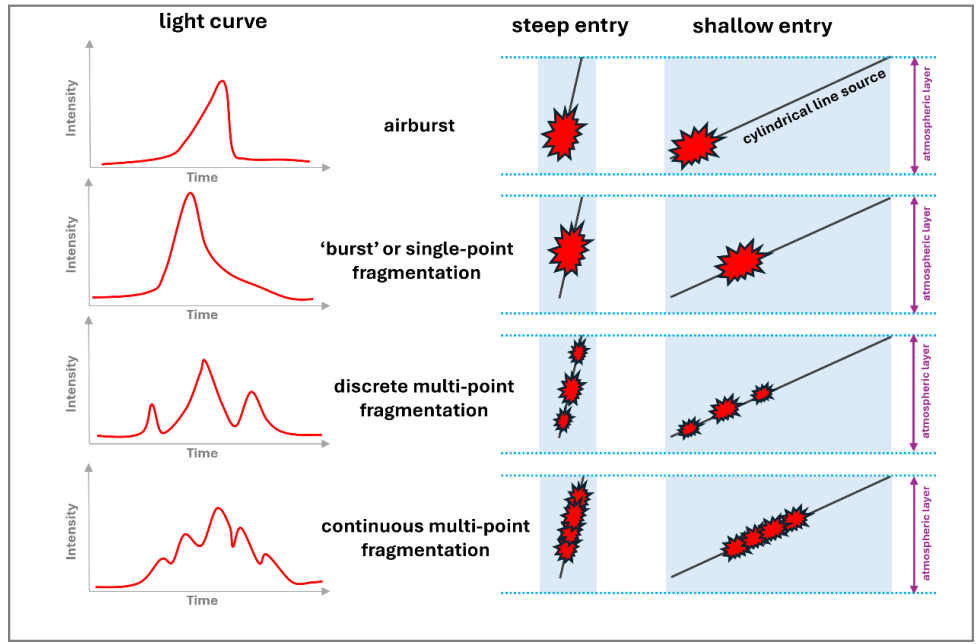

**Figure 2:** Idealized diagram, not drawn to scale and intended only as a rough conceptual representation, illustrates how a bolide's fragmentation behavior affects the resulting light curve. The left column shows four conceptual light curves, from a single-peaked airburst to continuous multi-point fragmentation, while the right panels depict how different geometry scenarios might look for steep versus shallow entries.

The bolide's associated light curve can be diagnostic in identifying these processes, as its shape, duration, and peak structure are sensitive to fragmentation events and can be leveraged to infer critical details about energy release and shock formation (Silber, 2024b; Silber and Brown, 2014; Wilson et al., 2025). **Figure 2** represents an idealized schematic illustrating how a bolide's fragmentation behavior can lead to characteristic light curve signatures. We omit a simple, non-fragmenting asteroid case from our diagram. On the left, four conceptual light curves are shown, each representing a distinct breakup pattern: an airburst with a single narrow peak, a single-point fragmentation with one broader peak, a discrete multi-point fragmentation with a series of peaks, and a continuous multi-point fragmentation producing multiple overlapping peaks. To the right, each fragmentation type is shown with both steep and shallow entry trajectories, emphasizing how the bolide's path through the atmosphere influences the spatial distribution of fragmentation episodes and the accompanying shock formation. In a steep entry, the energy is deposited within a relatively narrow vertical column, whereas a shallow entry spreads the disruption over a longer atmospheric path. These variations in breakup processes, combined with differences in entry angles, result in the diverse light curve shapes observed and offer clues about the shock generation and total energy release of the bolide event. Moreover, information gleaned from simultaneous infrasound and light curve analysis is of extraordinary importance for improving our understanding of the atmospheric entry of rare superbolides and assessing their potential to become hazardous projectiles (Trigo-Rodríguez, 2022).

Here, we investigate the impact of these processes in the period-yield relationships using the CNEOS bolide light curve data. One of the primary objectives in this study was to infer the mode of shock deposition through the analysis of light curves, as these serve as a diagnostic for various physical processes taking place during a bolide entry. We aim to investigate possible links between types of shock, altitudes of fragmentation episodes, and bolide parameters, and feed these into the energy fits to investigate the linkage. To evaluate the robustness of the period-yield relations, we apply a bootstrap model, a powerful statistical technique that repeatedly resamples the data to produce multiple 'bootstrapped' datasets (Efron, 1992; Efron and Tibshirani, 1993). Each resampling run estimates the parameters of interest (e.g., slope and intercept in period–energy fits), and the collection of those estimates is used to derive robust confidence intervals. The bootstrap approach offers a more comprehensive view of the variability and reliability of the derived coefficients than a single, static fit. In principle, testing the reliability of period-based energy relations under diverse event conditions will allow an assessment of the parameter ranges in which these relations are most reliable, or where they fall short.

## 2. Methods

### 2.1 Bolide Dataset

A bolide infrasound detections database, including measured signal parameters (see **Appendix B**), was compiled by drawing on multiple published sources that focused on bolides detected by U.S. government sensors and subsequently listed on the CNEOS website (Ens et al., 2012; Gi and Brown, 2017; Hupe et al., 2024; Ott et al., 2021; Ott et al., 2019; Pilger et al., 2020; Silber et al., 2024; Silber et al., 2011) (**Figure 3**). The CNEOS database supplies contextual metadata to reported energy yields, such as event time, location, and frequently a velocity vector and corresponding light-curve data (e.g., Silber, 2024b; Silber and Sawal, 2025). Where velocity vectors were available (95 events), we derived entry angles, azimuths, and orbital parameters based on the method of Peña-Asensio et al. (2022), which then informed about each bolide's physical properties.

All but 10 bolides had infrasound signals detected by the stations of the International Monitoring System (IMS) of the Preparatory Commission for the Comprehensive Nuclear-Test-Ban Treaty Organization (CTBTO PrepCom), from which signal periods were derived. To ensure consistency and verify data quality, we recalculated source-to-station distances, back azimuths (direction of infrasound wave arrival at a station), and celerities (defined here as the average acoustic propagation speed from source to receiver). We discarded any entries deemed outliers, most notably those with anomalously large back azimuth discrepancies (where the difference between observed and theoretical values is >50°) or unrealistically high celerities (>360 m/s). In keeping with established energy-relation derivations in prior studies, we also excluded signals that travelled through the thermospheric waveguide (celerity <220 m/s (Negraru et al., 2010)). Therefore, all events have celerities between 220 – 360 m/s and back azimuths deviations <50°.

This refinement process yielded 138 distinct bolide events and 362 total detections (**Figure 3**), constituting the largest single consolidated dataset of infrasound-detected CNEOS bolides to date. Among these, 69 bolides were detected by one station globally, whereas the remaining 69 were observed across two or more infrasound stations worldwide. The number of distinct stations that made detections was 59. Out of these, 53 are the IMS stations.

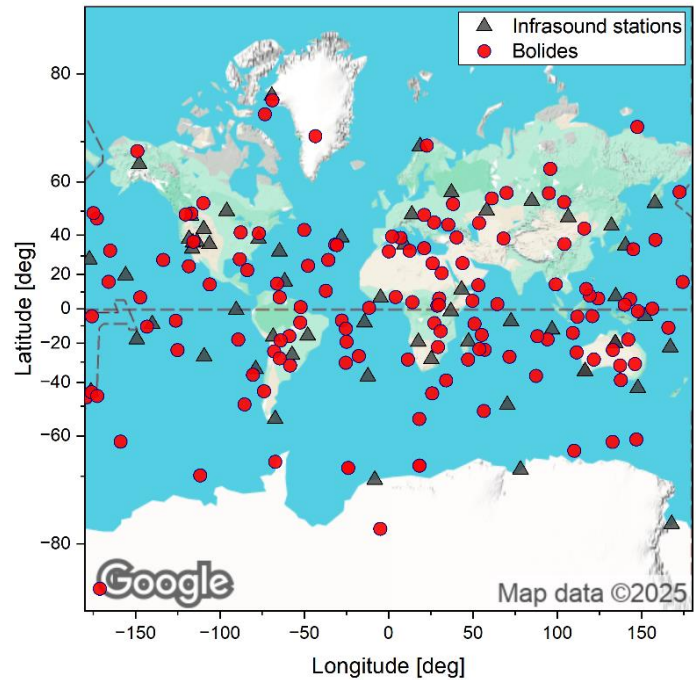

**Figure 3:** Global map showing all 138 bolides (red circles) investigated in this work and 59 distinct infrasound stations that recorded detections (grey triangles).

For these bolides, we retrieved event-specific atmospheric profiles from the Ground-2-Space (G2S) model hosted by University of Mississippi (Drob et al., 2003; Hetzer, 2024) along each source–station path. Earlier work has shown that Doppler shifts due to stratospheric winds can alter the observed period by more than ~10% (Ens et al., 2012). However, they also noted that uncertainties in wind velocity can be as large as the mean itself, and therefore did not correct for Doppler shift. In this study, we explore both raw and Doppler corrected measured periods. We make Doppler shift corrections following the approach outlined in ReVelle (2010). Since infrasound propagation is assumed to be dominated by the stratospheric waveguide, we consider the altitude span of 40–60 km in a range-dependent fashion, computing mean and standard deviation values of the wind field. We also accounted for temporal variations, which is particularly relevant for long-distance infrasound paths requiring multi-hour propagation.

**Figure 4** provides histograms illustrating the distributions of bolide parameters available in our dataset, including (top row) peak brightness altitude, source-to-station distance, and entry velocity, (middle row) total energy (on a log scale), entry angle, and impactor diameter, and (bottom row) dynamic strength (also log-scaled), impactor density, and the Tisserand parameter. The Tisserand parameter is a near-constant measure in the restricted three-body problem (e.g., Sun–planet–small body) that characterizes how a smaller object's orbit changes, or fails to change, before and after a close planetary encounter. In practical terms, the Tisserand parameter computed with respect to a given planet helps categorize interplanetary bodies (e.g., comets vs. asteroids) by capturing relationships among orbital elements such as semi-major axis, eccentricity, and inclination. For instance, many Jupiter-family comets have Tisserand values between 2 and 3 (Murray and Dermott, 1999). Not all events in the database have complete information for these parameters, which is reflected in some histograms having lower counts than others. Nonetheless, these plots demonstrate the range and diversity of bolide characteristics captured in our final dataset.

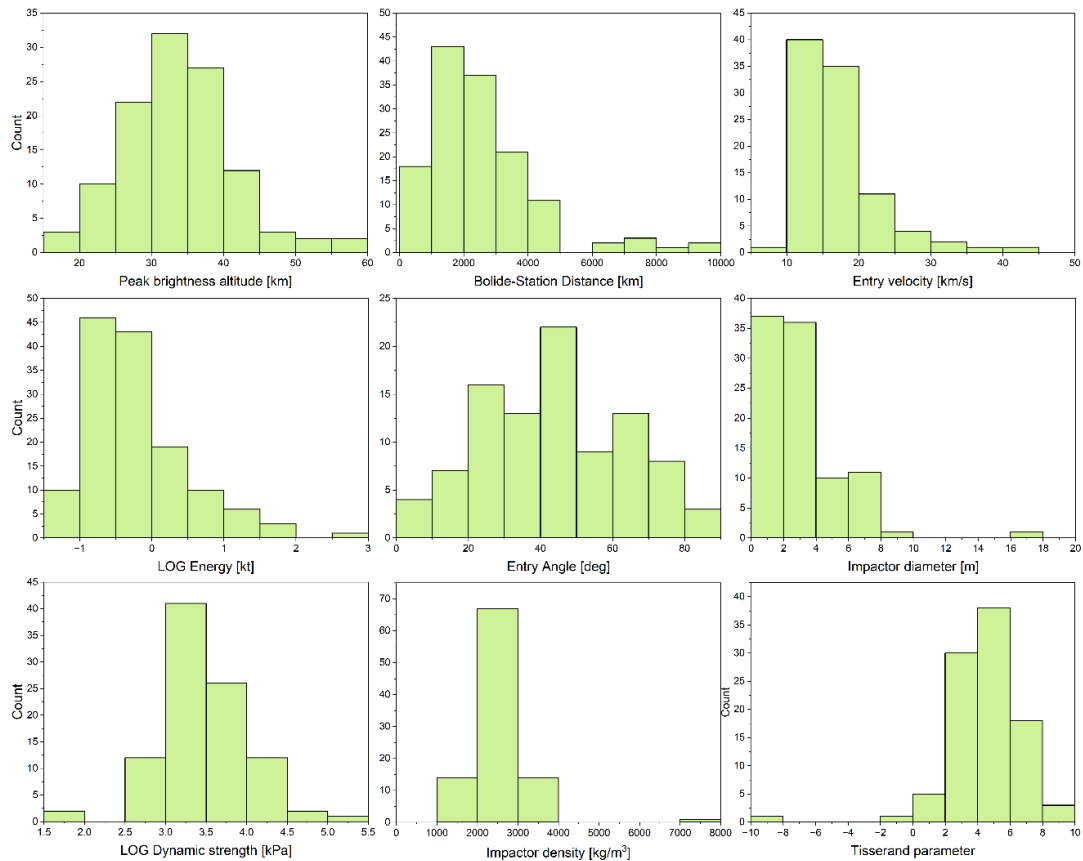

**Figure 4:** Histograms showing the distributions of bolide parameters available in our dataset: peak brightness altitude, source-to-station distance, entry velocity, total impact energy (on a log scale), entry angle, impactor diameter, dynamic strength (also log-scaled), impactor density, and the Tisserand parameter.

The CNEOS database hosts an inventory of light curves (LC) associated with most bolide events up to and including most of 2022. Presently, these come in a PDF format, requiring digitization. We gathered all available light curves, digitized and processed them following the analysis and classification algorithm developed by Silber and Sawal (2025). This algorithm uses the Savitzky-Golay filtering, prominence-based peak detection, and gradient analysis to automate the identification of fragmentation events and energy release modes. Silber and Sawal (2025) devised five classification schemes based on the light curve features to extract the dominant process of energy deposition (e.g., airburst, discrete fragmentation, continuous fragmentation). For the purposes of this study, we adopted a simplified light curve classification scheme: events with a single, dominant fragmentation episode (LC class 1) versus those exhibiting multiple fragmentation episodes, whether continuous or discrete (LC class 2). This simplified categorization enabled us to probe whether repeated fragmentation episodes (LC class 2) might broaden the infrasound signal relative to single-burst events (LC class 1), thereby clarifying fragmentation-driven influences on the measured period. Only three events did not have the light curve available.

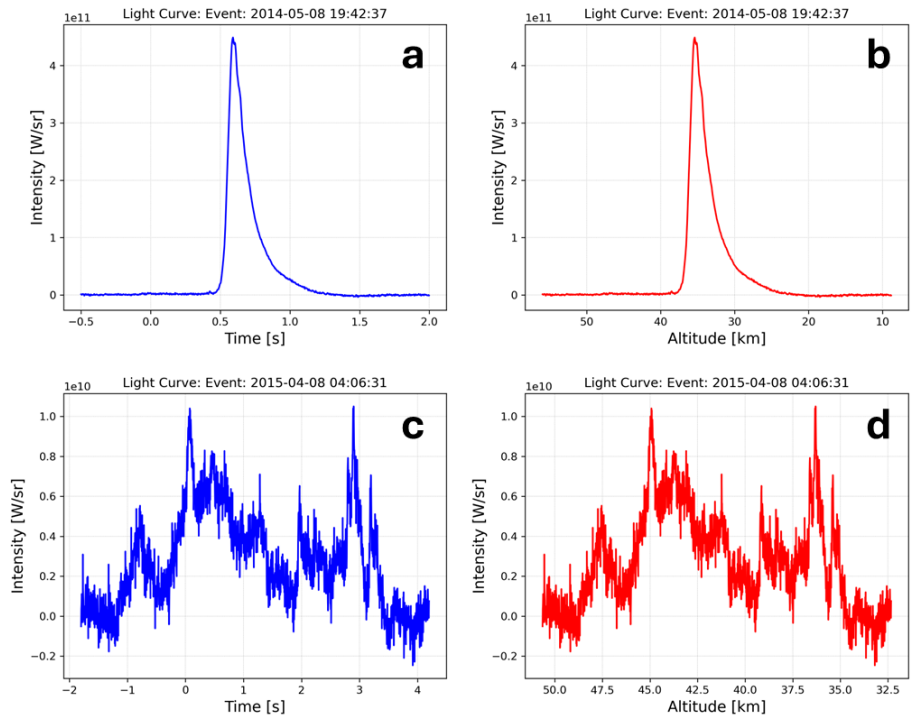

**Figure 5:** Representative examples of different classes of light curves, plotted as a function of the USG-provided time (left) and calculated altitude assuming no deceleration (right). Panels (a) and (b) show LC class 1, depicting a single dominant fragmentation event, where a single abrupt peak unfolds within 2 s and spans only about 4 km in altitude (37–33 km). This single fragmentation event is consistent with a classic 'airburst' in which the entire object is abruptly vaporized due to the fragmentation. Panels (c) and (d) show LC class 2. This is a multi-fragmentation event, with multiple bursts across 4 s and an altitude range from 48 km down to 35 km, indicative of energy deposition over 13 km.

We analyzed the light curves in two steps. First, we applied our classification algorithm to all available LCs to determine their respective types. Second, for CNEOS events that provided both a velocity vector and a peak brightness altitude, we anchored the altitude to the peak brightness point identified in the light curve and extrapolated the altitude range from there. Here, as the first approximation, we assume constant velocity. This approach allowed us to examine the onset or conclusion of fragmentation events more thoroughly, rather than focusing solely on the peak brightness altitude. A representative example of this step in our analysis is shown in **Figure 5**. Here, we compare two distinct bolide fragmentation scenarios by plotting light curve intensity (in W/sr) against both time and altitude. These examples demonstrate our approach to discerned multi-fragmentation versus a single, abrupt burst by analyzing the light curve shape and its corresponding altitude profile. As noted earlier, the fragmentation can be rather complex, where multiple fragmentation episodes can be discrete or continuous (see **Figure 2**). In the latter case, this could result in persistent emissions along the wake. However, we do not make that specific differentiation here, but rather focus on single burst versus multi-bursts, regardless of how they might have occurred.

**2.2 Ordinary Least Squares (OLS) Partitioning**

In their study, Ens et al. (2012) employed a multivariate approach that factored in parameters like range and amplitude; however, their analysis did not account for altitude or light curve classification, both of which can critically alter where and how shock energy is deposited along the bolide trajectory. Moreover, amplitude-based fits typically require explicit propagation corrections, making range an even more central parameter. In our study, by contrast, we focus on period rather than amplitude, which lessens the need for direct range corrections and enables us to isolate other physical factors (e.g., altitude, multi-episode fragmentation) that might influence the period–energy relationship.

Fundamentally, we adopt a log–log regression model of the form: $\log(P) = m\log(E) + c$. Recognizing that infrasound detections can be influenced by source altitude, entry angle, velocity, and similar parameters, we analyze not only the dataset as a whole but also partition it into subsets. For instance, we might isolate only events below a certain altitude threshold or within a specific velocity range; likewise, we can set distance ≤1000 km or altitude <20 km. Each such partition defines a dataset subset for which we perform a distinct ordinary least squares (OLS) fit. This step mitigates confounding effects (Montgomery et al., 2021) by preventing a single, global fit from obscuring systematic differences in event geometry or fragmentation type. Moreover, by initially running OLS on narrowly defined partitions and then applying a bootstrap framework, we can gauge more transparently how each parameter impacts the resulting slope and intercept. As in previous studies, we pursue two approaches in our analysis: (1) treating each station detection independently, and (2) computing a mean signal period for bolide events with multiple stations, by averaging the individual detections across those stations.

Within our database, each event at minimum includes the signal period, the total impact energy reported by CNEOS (in kt), and the source-to-station distance (in kilometers); in all but one case, a peak-brightness altitude (in kilometers) is also available. A subset of 135 bolides includes a light-curve classification, and 95 bolides contain additional parameters (entry angle, velocity, mass, density, diameter). Rows missing a specific parameter are excluded only from analyses involving that variable, thus preserving the maximum number of valid records for partitions that do not rely on it. For each partition (e.g., distance ≤1000 km, altitude <20 km, velocity <20 km/s, etc.), we perform fitting (Montgomery et al., 2021) using the *statsmodels.api.OLS* routine in Python, yielding a slope $\hat{m}$ and intercept $\hat{c}$. We also apply bin permutations of up to five parameters in the analysis. Since many published energy-period formulas use the inverted form: $\log(E) = A\log(P) + B$, we convert via $A = 1/\hat{m}$ and $B = 1/\hat{c}$, enabling direct comparisons with reference relations in the literature (e.g., Gi & Brown (2017), Ens et al. (2012), and ReVelle (1997)). A single OLS pass produces one pair $(\hat{m}, \hat{c})$ and hence one pair $(A, B)$. However, we also seek to characterize uncertainties, given that small sample sizes or physically diverse events can cause significant variance in slope estimates. Consequently, a bootstrap framework is employed to estimate confidence intervals for $(\hat{m}, \hat{c}, A, B)$, hereby assessing the reliability of each partition's fit under repeated sampling.

**2.3 Bootstrap Approach for Period–Energy Fitting**

To robustly estimate the variability and confidence intervals of our fitted slope/intercept pairs, and their inverted counterparts, we employ a bootstrap sampling framework *(Efron, 1981, 1992; Efron and Tibshirani, 1993; Hastie et al., 2009)*. For each partition of the dataset, we draw the same number of rows as in the original subset with replacement, repeatedly generating "pseudo-datasets." We then fit the OLS model $\log(P) = m_i \log(E) + c_i$ on each pseudo-dataset and invert the resulting slope and intercept. By iterating this process $n$ times ($n$ = 2000), we accumulate sets $\{m_i, c_i\}$ and $\left\{A_i = \frac{1}{m_i}, B_i = \frac{1}{c_i}\right\}$, corresponding to each bootstrap iteration $i$. From these distributions, we obtain not only the means $(\bar{A}, \bar{B})$ but also standard deviations and 95% confidence intervals.

A key point is that we transform each pair $(m_i, c_i)$ individually into $(A_i, B_i)$. This captures the true variability of the inverted parameters $\log(E) - \log(P)$, rather than merely inverting the mean slope $\bar{m}$. Indeed, $\bar{A} \neq 1/\hat{m}$ once distributions are non-linear. After $n$ bootstrap iterations, we collect distributions $\{m_i\}$, $\{c_i\}$, $\{A_i\}$, $\{B_i\}$. We then compute the mean values $\bar{m}$, $\bar{c}$, $\bar{A}$, $\bar{B}$, along with their standard deviations. Moreover, we derive confidence intervals in two ways: (i) normal approximation ($\bar{A} \pm z\sigma_A$ and $\bar{B} \pm z\sigma_B$ for 95% intervals, where $z$ =1.96), which is appropriate if $\{A_i\}$ and $\{B_i\}$ are roughly symmetric and (ii) percentile-based, where we take the empirical 2.5th and 97.5th percentiles of $\{A_i\}$ and $\{B_i\}$. This is often more robust if the distributions are skewed or if normality is not a good approximation. Comparing these two intervals helps us assess the degree of symmetry or skew in the underlying distributions.

To discern how specific parameters, such as source-to-station distance, peak brightness altitude, entry angle, impact velocity, or impactor properties, affect the period–energy relation, we subdivide the dataset into bins. For instance, distance bins might include ($d \leq 1000$, $\leq 3000$, $\leq 5000$, $\leq 10000$, all) while altitude bins include ($z_{pb}$ <20, <30, <40, all). We then run separate bootstrap-based OLS fits for each bin (or bin combination), storing: (1) the number of data points $N$ in that subset, (2) the mean $\bar{m}$, $\bar{c}$, $\bar{A}$, $\bar{B}$ with their standard deviations and confidence intervals, and (3) a diagnostic figure comparing the resulting line to existing references. A similar procedure applies to partitions defined by entry angle (<30°, 30°-60°, >60°), mass bins, density ranges, velocity, diameter thresholds, and light curve class classifications. This partition-based, multi-parameter nested approach systematically tests whether, for instance, shallow-entry vs. steep-entry bolides or high-altitude vs. low-altitude bursts exhibit distinct slope–intercept behaviors. By combining partitioning with a bootstrap framework, we obtain a clearer, more statistically robust picture of how each subset's period–energy fit varies, and the degree of confidence we can place in that variation.

# 3. Results

## 3.1 Ordinary Least Squares Results

**Table 2** presents the results from the OLS analysis for different subset of bolides. $N$ is the number of subset samples. The fit parameters $A$ and $B$ are given from both uncorrected and Doppler-wind corrected data. The classical regression coefficient is given also. In single detections, the coefficients $A$ and $B$ across 31 primary populations, binned by up to 2 parameters, were 3.47±0.49 and -1.88±0.24, respectively. Specifically, the range was 2.60 ≤ $A$ ≤ 5.03, and -2.26 ≤ $B$ ≤ -1.18. While the absolute differences in the regression coefficients $A$ and $B$ may seem modest in numerical terms, they reflect statistically and physically significant deviations among partitioned populations.

**Table 2:** Summary of regression coefficients *A* and *B* derived through OLS across all detections.

| Distance [km] | N | A | B | r² | Doppler corrected A | B | r² |
|---|---|---|---|---|---|---|---|
| **global** | **362** | **3.29** | **-1.88** | **0.59** | 3.28 | -1.88 | 0.58 |
| d=all, z<30 km | 163 | 2.88 | -1.52 | 0.67 | 2.87 | -1.52 | 0.67 |
| d=all, z≤40 km | 282 | 3.15 | -1.77 | 0.63 | 3.13 | -1.76 | 0.62 |
| d≤1000 km, z<30 km | 11 | 3.10 | -1.60 | 0.66 | 2.99 | -1.55 | 0.66 |
| d≤1000 km, z≤40 km | 27 | 3.23 | -1.70 | 0.55 | 3.18 | -1.70 | 0.52 |
| d≤1000 km, z=all km | 37 | 3.60 | -1.89 | 0.47 | 3.60 | -1.92 | 0.43 |
| d≤3000 km, z<30 km | 63 | 3.55 | -1.78 | 0.57 | 3.45 | -1.72 | 0.58 |
| d≤3000 km, z≤40 km | 141 | 3.88 | -2.04 | 0.47 | 3.82 | -2.01 | 0.46 |
| d≤3000 km, z=all | 198 | 4.16 | -2.22 | 0.38 | 4.11 | -2.19 | 0.37 |
| d≤5000 km, z<30 km | 95 | 3.24 | -1.68 | 0.60 | 3.11 | -1.61 | 0.61 |
| d≤5000 km, z≤40 km | 206 | 3.59 | -1.96 | 0.50 | 3.50 | -1.92 | 0.50 |
| d≤5000 km, z=all | 279 | 3.82 | -2.11 | 0.43 | 3.74 | -2.07 | 0.42 |
| d≤10000 km, z<30 km | 143 | 3.01 | -1.57 | 0.65 | 2.98 | -1.56 | 0.65 |
| d≤10000 km, z≤40 km | 260 | 3.32 | -1.84 | 0.60 | 3.29 | -1.83 | 0.59 |
| d≤10000 km, z=all | 340 | 3.49 | -1.97 | 0.54 | 3.46 | -1.95 | 0.53 |
| θ<30°, v≤ 20 km/s | 69 | 3.00 | -1.81 | 0.72 | 2.96 | -1.76 | 0.71 |
| θ<30°, v=all | 84 | 2.95 | -1.69 | 0.72 | 2.92 | -1.65 | 0.72 |
| 30°≤θ<60°, v< 20 km/s | 70 | 4.09 | -2.15 | 0.33 | 3.92 | -2.15 | 0.30 |
| 30°≤θ<60°, v=all | 88 | 4.43 | -2.23 | 0.31 | 4.32 | -2.25 | 0.28 |
| θ≥60°, v≤ 20 km/s | 101 | 3.59 | -2.10 | 0.57 | 3.57 | -2.07 | 0.57 |
| θ≥60°, v=all | 107 | 3.61 | -2.11 | 0.57 | 3.57 | -2.07 | 0.56 |
| θ=all, v≤ 20 km/s | 240 | 3.25 | -1.86 | 0.65 | 3.26 | -1.88 | 0.62 |
| θ=all, > 20 km/s | 39 | 5.03 | -2.26 | 0.27 | 4.91 | -2.17 | 0.30 |
| 1e5≤mass<1e8 | 134 | 3.00 | -1.65 | 0.44 | 3.05 | -1.72 | 0.42 |
| 2000≤dens<3500 | 171 | 3.83 | -2.19 | 0.38 | 3.77 | -2.18 | 0.37 |
| dens≥3500 | 83 | 2.60 | -1.18 | 0.69 | 2.64 | -1.22 | 0.71 |
| 1≤diam<10 | 260 | 3.64 | -2.04 | 0.51 | 3.64 | -2.05 | 0.49 |
| diam=all | 277 | 3.27 | -1.83 | 0.62 | 3.27 | -1.84 | 0.61 |
| LC=1 | 156 | 3.31 | -1.92 | 0.46 | 3.26 | -1.86 | 0.46 |
| LC=2 | 203 | 3.33 | -1.88 | 0.59 | 3.30 | -1.88 | 0.58 |
| LC=all | 359 | 3.29 | -1.88 | 0.59 | 3.28 | -1.87 | 0.58 |

**Table 3:** Summary of regression coefficients *A* and *B* derived through OLS across averaged period events.

| Distance [km] | N | A | B | r² | Doppler corrected A | B | r² |
|---|---|---|---|---|---|---|---|
| **global** | **138** | **3.65** | **-2.05** | **0.45** | 3.60 | -2.01 | 0.46 |
| d=all, z<30 km | 35 | 3.07 | -1.60 | 0.71 | 3.05 | -1.57 | 0.71 |
| d=all, z≤40 km | 94 | 3.31 | -1.78 | 0.55 | 3.27 | -1.74 | 0.55 |
| θ<30°, v≤ 20 km/s | 19 | 3.80 | -2.12 | 0.54 | 3.78 | -2.10 | 0.54 |
| θ<30°, v=all | 27 | 3.50 | -1.86 | 0.59 | 3.51 | -1.86 | 0.59 |
| 30°≤θ<60°, v< 20 km/s | 36 | 4.11 | -2.05 | 0.37 | 4.05 | -2.00 | 0.37 |
| 30°≤θ<60°, v=all | 44 | 4.21 | -2.04 | 0.37 | 4.12 | -1.99 | 0.37 |
| θ≥60°, v≤ 20 km/s | 21 | 3.53 | -1.96 | 0.70 | 3.51 | -1.96 | 0.68 |
| θ≥60°, v=all | 24 | 3.67 | -2.10 | 0.62 | 3.65 | -2.09 | 0.60 |
| θ=all, v≤ 20 km/s | 76 | 3.66 | -1.96 | 0.56 | 3.62 | -1.93 | 0.55 |
| θ=all, > 20 km/s | 19 | 3.69 | -1.82 | 0.43 | 3.69 | -1.82 | 0.44 |
| 1e5≤mass<1e8 | 19 | 3.14 | -1.69 | 0.44 | 3.13 | -1.68 | 0.43 |
| 2000≤dens<3500 | 67 | 4.08 | -2.16 | 0.36 | 3.99 | -2.10 | 0.36 |
| dens≥3500 | 15 | 2.85 | -1.34 | 0.77 | 2.84 | -1.33 | 0.77 |
| 1≤diam<10 | 93 | 4.10 | -2.16 | 0.44 | 4.05 | -2.12 | 0.44 |
| diam=all | 96 | 3.71 | -1.97 | 0.52 | 3.67 | -1.94 | 0.52 |
| LC=1 | 41 | 3.63 | -2.13 | 0.40 | 3.61 | -2.11 | 0.40 |
| LC=2 | 94 | 3.95 | -2.12 | 0.41 | 3.91 | -2.09 | 0.41 |
| LC=all | 135 | 3.69 | -2.05 | 0.45 | 3.66 | -2.03 | 0.45 |

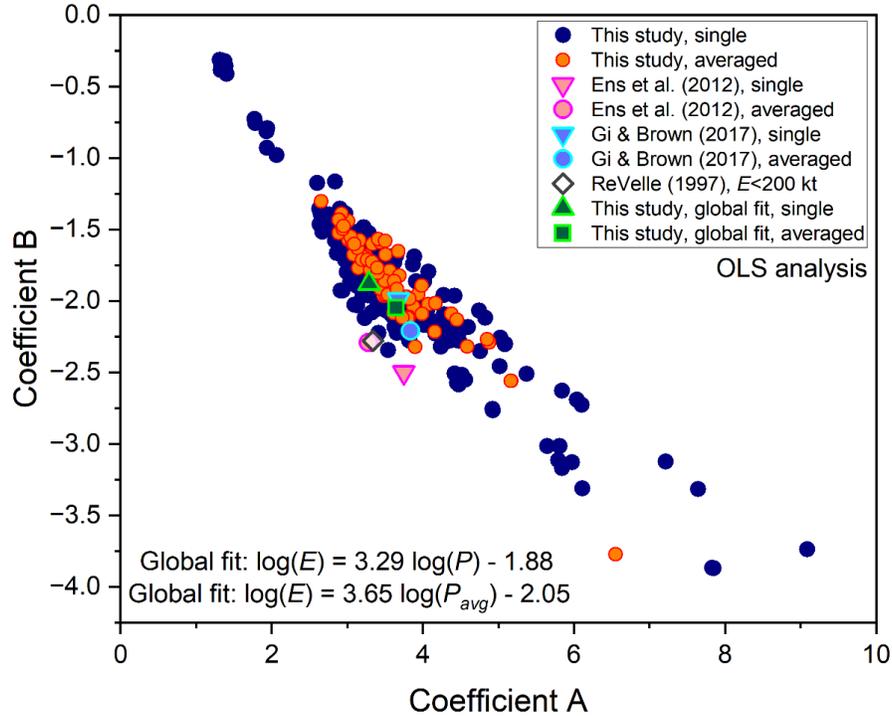

**Figure 6:** Distributions of OLS-derived coefficients *A* and *B* for both single-station (blue points) and multi-station averaged (orange points) period fits.

**Table 3** presents the same data but for averaged period detections. The coefficients *A* and *B* across 18 primary populations, binned by up to 2 parameters, were 3.63±0.37 and -1.92±0.22, respectively. Specifically, the range was 2.85 ≤ *A* ≤ 4.21, and −2.18 ≤ *B* ≤ -1.34.

**Figure 6** illustrates the distribution of best-fit coefficients *A* and *B* obtained from our OLS regressions across a range of bolide subsets. We performed a total of 509 fits for single-station detections, each corresponding to permutations of up to five parameters (e.g., altitude, distance, velocity, fragmentation type, light curve class), and 126 fits for multi-station averages (where distance was excluded). After removing any partition with insufficient data (*N*<20 in single detections, *N*<10 in averaged periods), we retained 274 and 70 pairs of coefficients, respectively.

Visually, the single-station data (blue points) cluster in a diagonal pattern, reflecting the fact that more negative intercepts (*B*) occur at higher slopes (*A*). The orange points from multi-station averages appear more tightly grouped, reflecting the often more stable period estimates when multiple stations contribute to a single, averaged period. These results emphasize both the overall trend and the variability inherent in subdividing bolide data by altitude, distance, and other parameters. We also overlaid coefficients from earlier studies alongside our best fits. Across all single-station and averaged period detections, respectively, our global regression yields:

$$\log(E) = 3.29 \log(P) - 1.88 \qquad (2),$$

$$\log(E) = 3.65 \log(\bar{P}) - 2.05 \qquad (3).$$

Some bins suggest a strong correlation between period and energy when the shock is observed closer to ground level and less 'smeared' by propagation. Tails at the lower end of the slope distribution arise in scenarios where energy release is more spread out, either by shallow geometry or multi-fragmentation combined with distant station observation.

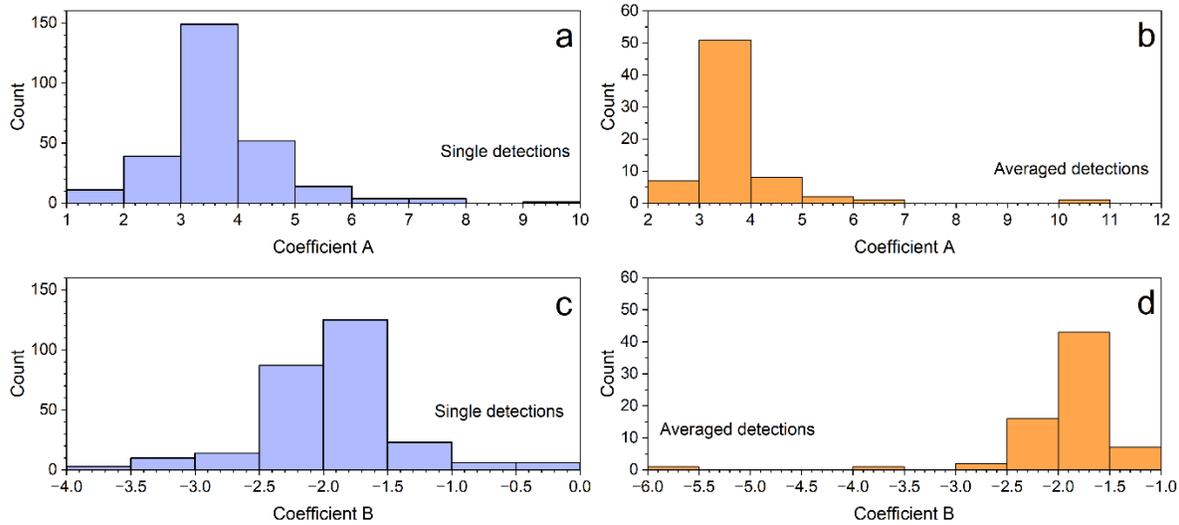

**Figure 7:** Histograms showing the spread in coefficients A and B as a function of populations of single and averaged detections. These include all our fits after outliers were removed.

**Figure 7** illustrates the overall variability and spread across all results in the form of histograms, subdivided by single (**Figures 7a and 7c**) and averaged populations (**Figures 7b and 7d**), and *A* and *B* coefficients.

Since our data set is substantially larger than in previous studies, we also performed the fitting using Doppler-corrected periods (ReVelle, 1974). Across all single-station and averaged period detections, respectively, our global regression yields:

$$\log(E) = 3.28 \log(P) - 1.88 \quad \text{(Doppler corrected periods)} \quad (4),$$

$$\log(E) = 3.60 \log(\overline{P}) - 2.01 \quad \text{(Doppler corrected periods)} \quad (5).$$

Compared to the non-corrected data, the resulting coefficients differed by only about 1.2% on average, with a maximum deviation of 4.3% in both coefficients *A* and *B*. This outcome is consistent with Ens et al. (2012), who similarly concluded that Doppler corrections have little effect when deriving period-based relations for a global population. Nevertheless, Doppler effects may still be important for individual-event analyses, as the averaging over numerous events can mask or dilute any single-event corrections.

To illustrate this point, we plotted the effect of Doppler shifts on measured infrasound periods (**Figure 8**). **Figure 8a** shows wind velocities along each source–station path for which wind data was available, with the shaded region highlighting distances of up to 6000 km where wind speeds often exceed ±50 m/s. **Figure 8b** compares the Doppler-corrected and raw measured periods, demonstrating that most points cluster near the 1:1 line, although some events deviate substantially at longer periods. The absolute difference between the Doppler-corrected and raw periods (**Figure 8c**) reveals that corrections can shift the period by several seconds, especially for greater source–station separations or higher wind speeds. **Figure 8d** 'zooms in' to 14 s, and more closely shows that most points lie near the 1:1 line.

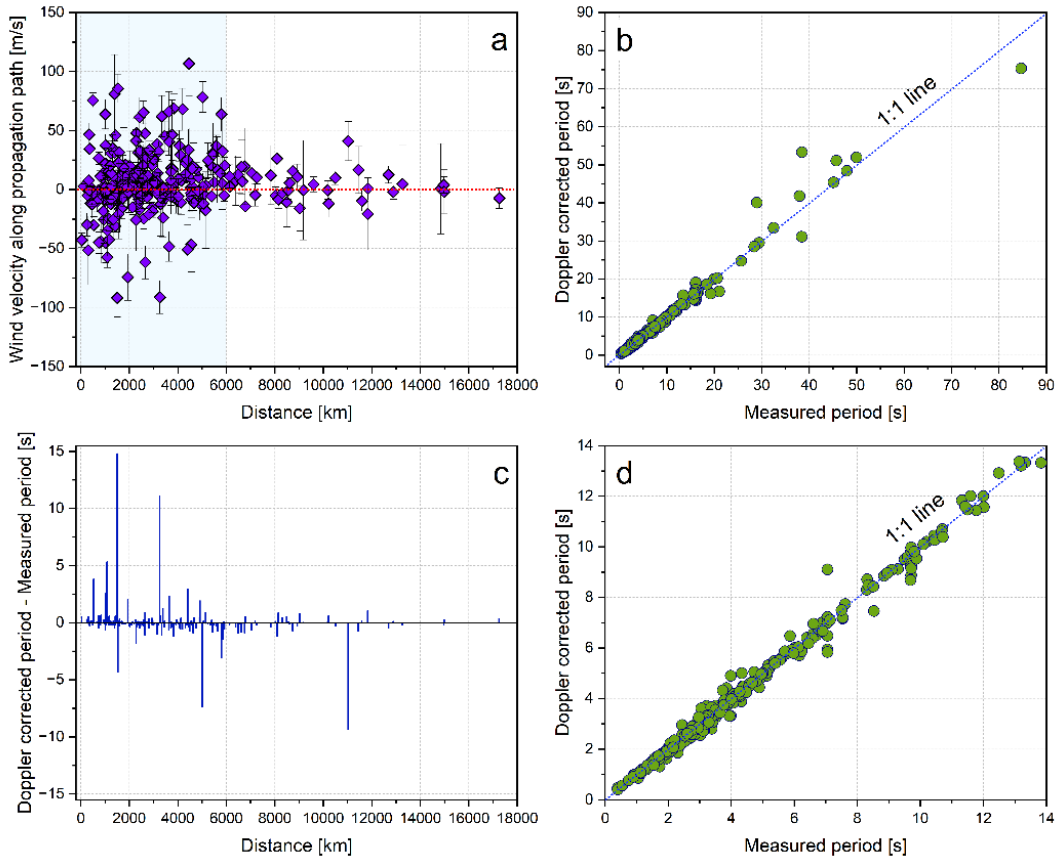

**Figure 8:** (a) Wind velocity along the source–station path, plotted against propagation distance. (b) Doppler-corrected versus measured period. (c) Difference between Doppler-corrected and measured periods, showing distance-dependent variations. (d) Zoomed-in version of (b) for shorter periods, again showing how stratospheric winds can alter observed infrasound periods.

### *3.2 Bootstrap Approach Results*

Using a bootstrap approach, we derived *A* and *B* coefficient pairs with their respective standard deviations and confidence bounds. Similar to the OLS analysis performed prior to this, for single-station detections, we included source-to-station distance as a factor, whereas for multi-station average periods we excluded it. Alongside global *A* and *B* pairs, we produced over 500 subset fits, each reflecting permutations of up to five parameters in a given subgroup. Before examining the correlations, we discarded any results with *N*<20 for single detections and *N*<10 for averaged detections. Further to this, we examined coefficient standard deviations and removed poor fits (i.e., large standard deviations), thereby ensuring only robust regressions were carried forward in the analysis. We were left with 250 and 50 pairs of coefficients *A* and *B*, respectively, for single detections and averaged detections. The global fits with uncertainties are as follows:

$$\log(E) = 3.30\,(\pm 0.14)\log(P) - 1.89(\pm 0.10) \qquad (6),$$

$$\log(E) = 3.71(\pm 0.42)\log(\overline{P}) - 2.07(\pm 0.21) \qquad (7).$$

Doppler corrected periods yield coefficients *A* and *B* that closely match those in Eq. (6) and Eq. (7), remaining well within their respective uncertainties.

**Figure 9** shows the coefficient *A* and *B* pairs for single (navy blue circles) and averaged detections (diamonds). We opted against plotting the error bars as they obscure the plot. For context, we also plotted single detections derived through the OLS method (light green circles). The global fits with error bars are shown with a triangle (Eq. (6)) and square (Eq. (7)).

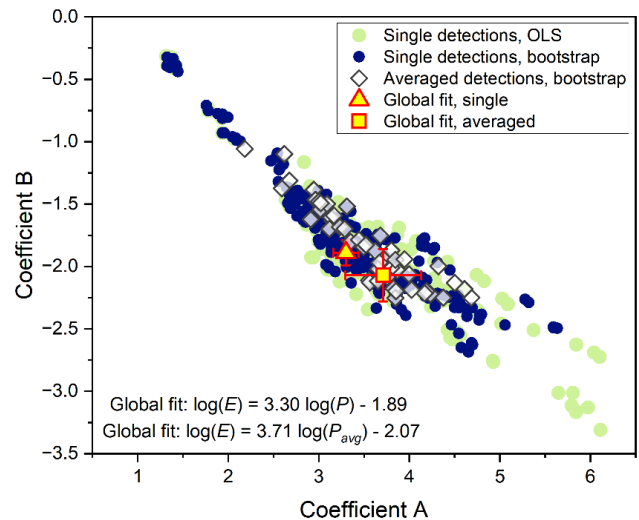

**Figure 9:** Coefficient *A* and *B* pairs for single detections (navy blue circles) and averaged detections (diamonds). Single detections derived with ordinary least squares (light green circles) are also shown, but some extreme points are cut off in this zoomed in version. Error bars for individual points are omitted to maintain clarity. The global fits, with error bars, appear as a triangle (Eq. (4)) and a square (Eq. (5)).

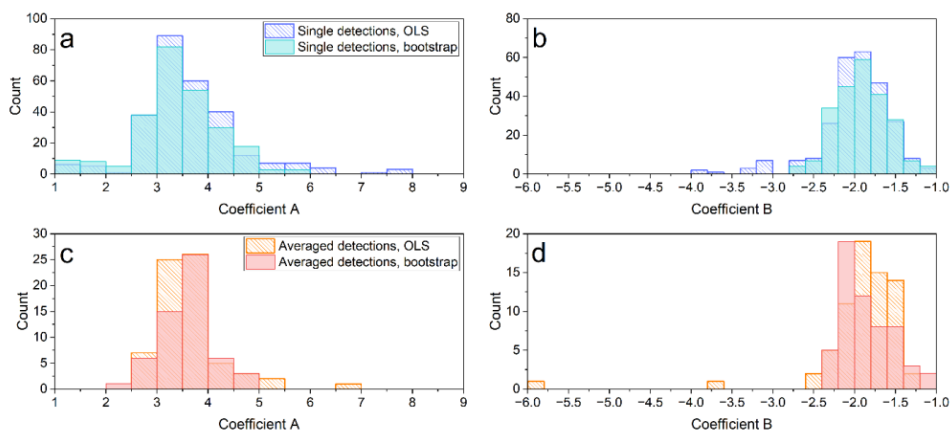

**Figure 10:** Histograms of coefficients A and B for single detections (panels a, b) and averaged detections (panels c, d). The ordinary least squares results are shown with hatched bars, while the bootstrap results are shown with solid bars. Note that the axes scale is different in all panels.

For additional context and to compare the outcomes from the two fitting approaches, we plotted the distributions of coefficients *A* and *B* under both OLS (hatched bars) and bootstrap (solid bars) approaches (**Figure 10**). **Figure 10a-b** displays histograms of coefficients *A* and *B* for single detections and **Figure 10c-d** for averaged detections. This allows a straightforward visual comparison of how each method influences the coefficient estimates for single versus averaged detections.

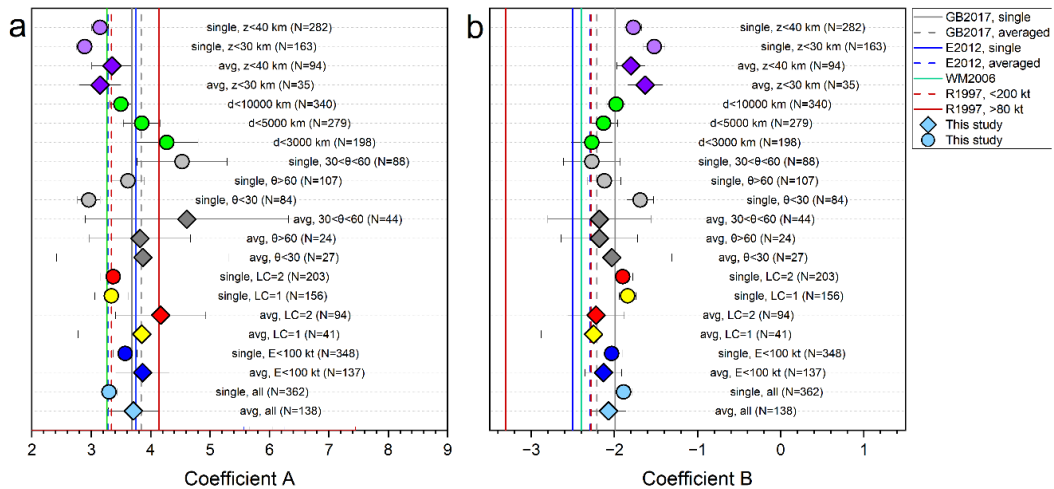

**Figure 11:** Coefficients A (a) and B (b) from $\log(E) = A\log(P) + B$, plotted for numerous bolide subgroups. Each point represents a distinct partition (e.g., by altitude, angle, distance, fragmentation type), with error bars indicating fit uncertainties. Reference lines from Gi & Brown (2017), Ens et al. (2012), Whitaker and Mutschlecner (2006), Whitaker (2023), and ReVelle (1997) are shown for comparison, demonstrating how geometry (e.g., shallow vs. steep angles) and multi-fragmentation (LC class 1 vs. 2) can shift slopes and intercepts away from classic global values. The circles and diamonds represent fits for single station and averaged events, respectively.

**Figure 11** provides an at-a-glance summary of the coefficient *A* and *B* for various subsets of single and averaged detections. Each label indicates a subset's primary parameter (e.g., altitude, distance, entry angle, light curve class, or yield range), and the corresponding sample size (*N*). The figure also shows published coefficients from Gi and Brown (2017), Ens et al. (2012), and ReVelle (1997), plotted as vertical lines for *A* and reference lines for *B*, to place the new results in context. This layout allows a straightforward comparison of how each subset's best-fit slope (**Figure 11a**) and intercept (**Figure 11b**) compares to previously reported values, motivating an evaluation of how parameters such as altitude, range, or fragmentation behavior might influence the period–energy relationship.

In **Figure 12**, we show the bolide period-yield regression fits obtained through this study and compared to all earlier studies (**Figure 12a**). Single detection fits and averaged fits are shown in **Figures 12b** and **12c**, in the context of relevant earlier studies. Finally, shown in **Figure 12d** are the global fits derived through bootstrap method alongside fits for LC1 and LC2 populations.

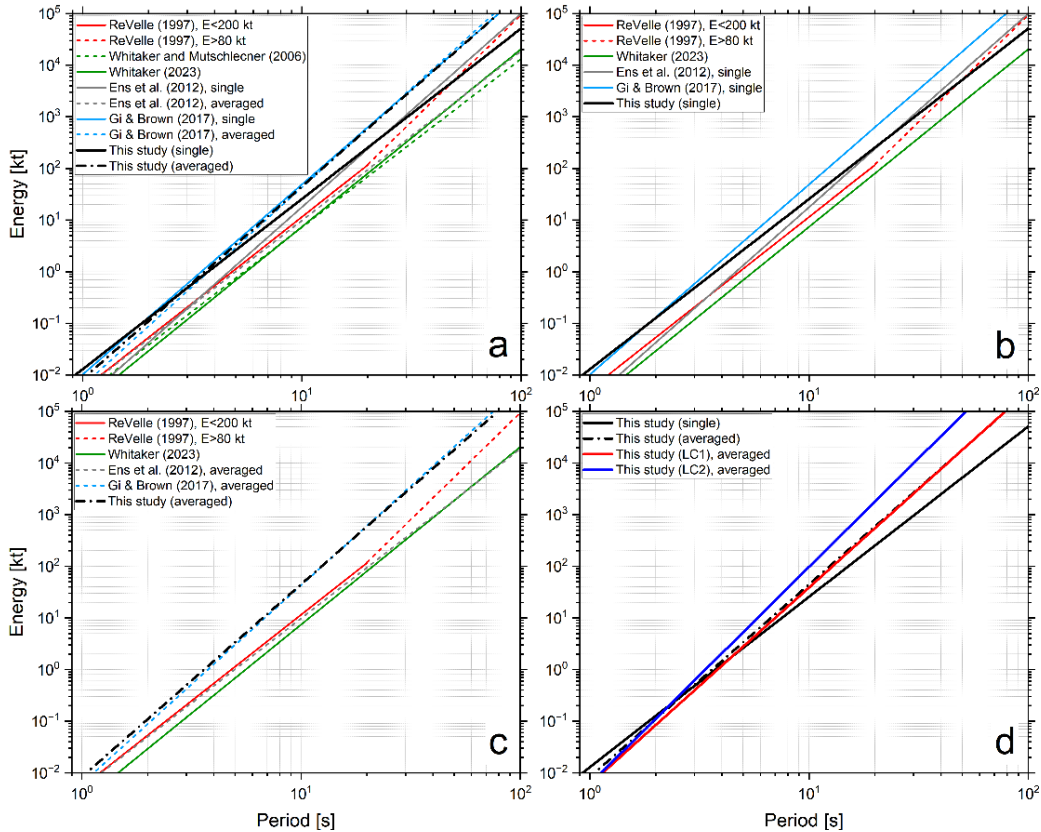

**Figure 12:** Log–log plot of total impact energy versus signal period comparing (a) published period–energy relations and those derived in this study. Fits for ingle station detections (b) and averaged periods (c) are plotted separately alongside earlier studies. In panel (d), we show the global fits alongside LC1 and LC2 (averaged).

### 3.3 Period and Yield versus Multi-Parameter Variables

We computed Spearman correlation (Spearman, 1904) matrices to examine how primary parameters relate to one another across our bolide dataset. The parameters are period, CNEOS energy, impact velocity, entry angle, altitude of peak brightness, meteoroid density, meteoroid diameter, and meteoroid dynamical strength. The Spearman's rank correlation coefficient is a nonparametric measure of how two variables co-vary in a monotonic manner, using rank orders rather than raw values. By forgoing assumptions of linearity and normality, it remains robust in the presence of outliers or skewed data, capturing whether one variable consistently increases or decreases with the other. **Figure 13** presents correlation matrices for averaged period bolides. **Figure 13a** includes the entire dataset, while **Figures 13b** and **13c** show two primary fragmentation classes, LC class 1 (single-burst or airburst events) and LC class 2 (multi-fragmentation). Each cell in the matrices is color-coded according to the strength and sign of the correlation, with red indicating a positive correlation and blue a negative correlation.

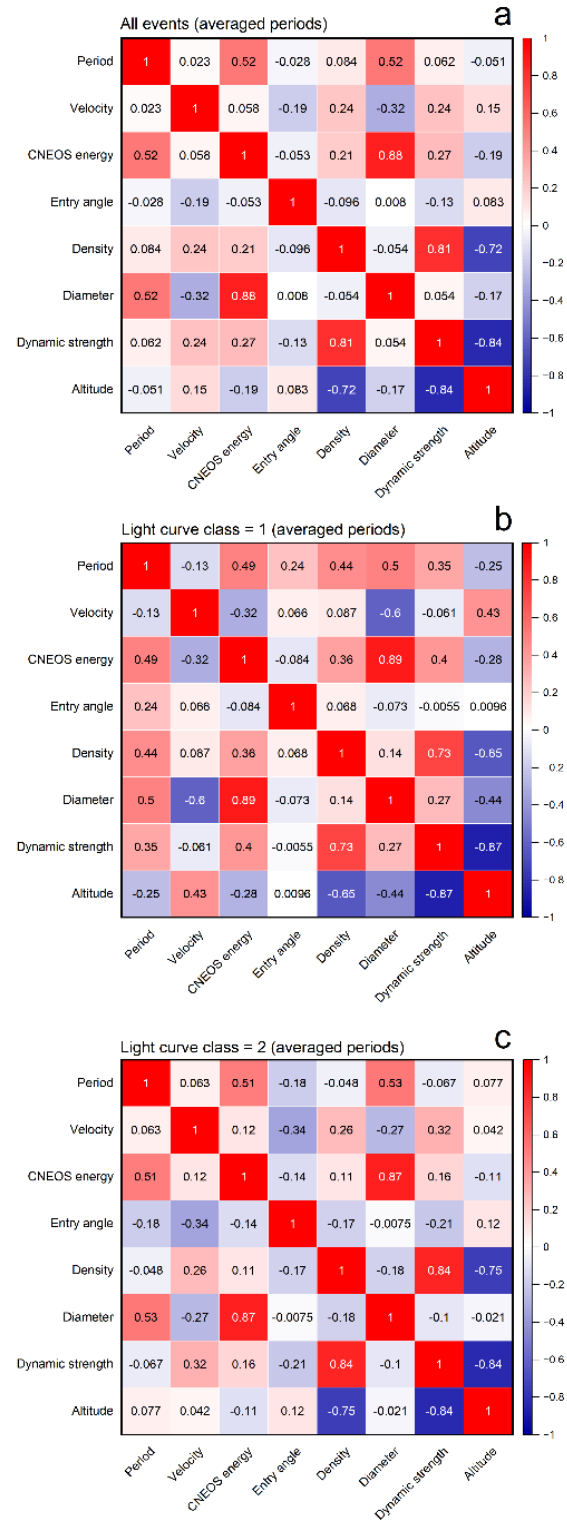

**Figure 13:** Correlation matrices for averaged-period bolides across (a) all events, (b) single-burst (LC class 1) events, and (c) multi-burst (LC class 2) events. Each panel shows pairwise Spearman correlation coefficients among various parameters: infrasound period, CNEOS energy, entry angle, velocity, diameter, density, dynamic strength, and peak-brightness altitude.

As expected, the highest correlation coefficient is between CNEOS energy and meteoroid diameter. The meteoroid diameter is typically estimated from the reported yield. However, we also see a strong correlation when the diameter is independently measured from cosmogenic nuclides in recovered meteorites (e.g., Jenniskens et al., 2021; Jenniskens et al., 2009). Another strong correlation exists between dynamical strength and meteoroid density, both parameters are related to the atmospheric altitude at which key phenomena are observed. CNEOS energy and period are less correlated, reflecting the influence of other parameters in the relation.

To explore the relationship between bolide energy and the maximum distance at which the infrasound signal was detected, we plotted the detection range as a function of satellite-estimated yield (**Figure 14**). Examining this trend provides a straightforward means of assessing how far infrasound from a given bolide might travel before dropping below detectability thresholds. As in previous work (e.g., Ens et al., 2012), an empirical upper-range curve can be superimposed to represent a typical outer envelope of observed detections, thereby highlighting how wind conditions or other propagation effects can extend (or limit) the observable range for lower-energy events. This approach also offers a simple visualization for distinguishing bolide signals from other atmospheric sources by illustrating whether an observed detection range is consistent with the event's reported yield. Our expression for detection range, with energy in units of kt and distance in units of km, is:

$$\log(R) = 0.28\log(E) + 3.85 \qquad\qquad (6).$$

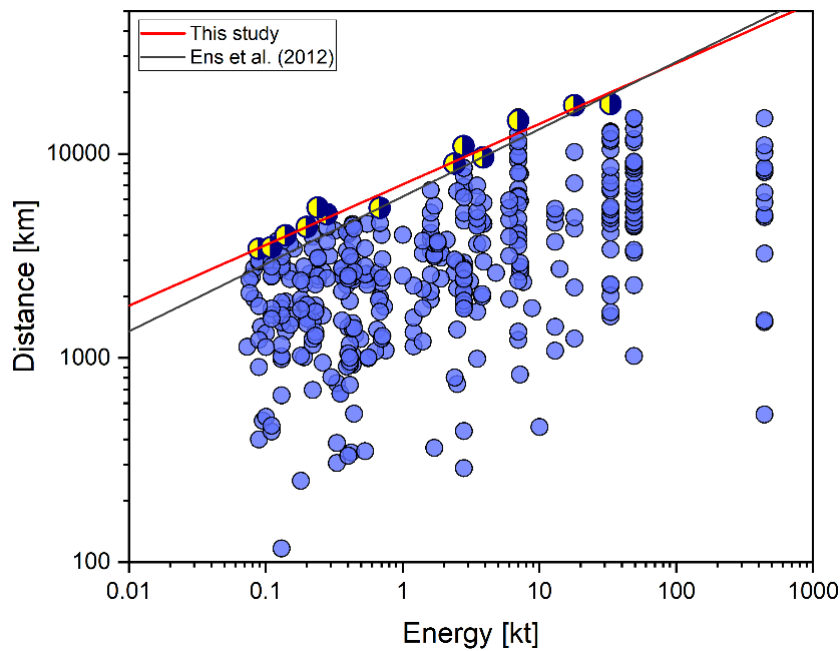

**Figure 14:** Detection range (km) versus total impact energy (kt) for all infrasound stations that recorded a positive signal. The red line represents the upper-range envelope (yellow-black circles) derived in this study, while the black line shows the relation from Ens et al. (2012), $\log(R) = 0.33\log(E) + 3.79$. Each blue circle corresponds to a unique detection, illustrating how measured ranges vary with event yield; the superposed curves provide a simple upper bound on the distances over which infrasound from bolides of given energies can be detected.

## 4. Discussion

### 4.1 OLS vs. Bootstrap and Value of Sub-Group Binning

OLS provides a straightforward point estimate of the best-fit line under classic linear regression assumptions, while bootstrap resampling quantifies how sensitive those OLS results are to variations in the underlying data. The close agreement between the OLS (**Figures 6 and 7**) and bootstrap-derived results (**Figures 9 and 10**) reinforces the stability of our core period–energy relationships, demonstrating that removing or resampling small subsets of events does not drastically alter the estimated slopes ($A$) or intercepts ($B$). The bootstrap analysis further indicates where outlier events or small partitions (e.g., bins characterized by shallow angles and low altitudes) can yield slope–intercept pairs that can deviate from the 'typical' 3–4 slope region and −2 intercept. While OLS averages over these extremes, the bootstrap approach (see tails in **Figures 9 and 10**) illuminates their frequency and variability, providing robust confidence measures for each subgroup fit. By applying multi-parameter binning, using both OLS and bootstrap, we identify distinct phenomena like steeper slopes at low altitudes and short ranges, and shallower slopes at shallow angles or distant stations. Our findings (as depicted in **Figures 9 and 10**) confirm that these variations are physically meaningful and represent distinct shock-physics regimes, thereby supporting the need for subgroup-specific fits rather than relying on a single universal formula.

### 4.2 Sensitivity of Coefficients A and B

Our expanded dataset exhibits significant variability in coefficients $A$ (slope) and $B$ (intercept), reflecting the complex interplay of geometry, fragmentation type, and propagation effects in period–energy relationships. Although our global fits (Eq. (2-5) generally conform to earlier literature (e.g., ReVelle, 1997; Ens et al., 2012; Gi & Brown, 2017), partitioning the data by altitude, angle, distance, or multi-fragmentation yields values of $A$ that may exceed 4.0 or fall to 2.5–3.0, suggesting a wider parameter space than that provided by a single global fit. These variations can be attributed to shock-physics principles. Deeper or steeper trajectories yield stronger coupling between energy and the dominant period, whereas shallower entries or large station distances dilute period sensitivity to energy. Equally important, multi-fragmentation (LC class 2) tend to amplify period changes relative to single bursts (LC class 1) because overlapping bursts (i.e., increased energy deposition through fragmentation episodes), or superposition of higher frequency peaks reinforce low-frequency signal components.

Considerable scatter is observed in both the original datasets (often nuclear test signals) and the subsequent bolide analyses (e.g., Mas-Sanz et al., 2020; Silber, 2024a). Such scatter might arise for reasons spanning signals emanating from different parts of the trajectory (Silber, 2024a; Silber et al., 2009), altitude effects (Edwards et al., 2006), source-to-station range effects (Gi and Brown, 2017), station geometry (Pilger et al., 2015), station noise (Bowman et al., 2005), wind-driven Doppler shifts (Ens et al., 2012), and measurement uncertainties (Golden and Negraru, 2011). Though infrasound signal period, compared to signal amplitude, is less sensitive to range, it might still exhibit variations as stations may receive signals from different portions of the bolide path (Silber and Bowman, 2025; Silber, 2025). Furthermore, the range of shock configurations (from straightforward cylindrical line sources to localized quasi-spherical explosions) might plausibly explain the significant scatter in simple period–energy relations. Thus, determining the prevailing shock geometry (cylindrical vs.

quasi-spherical) is critical for accurately linking measured infrasound periods to underlying energy deposition physics. We further discuss these points in the next sections.

### 4.3 Light-Curve Classification and the Fragmentation Process

Despite their practical utility in distinguishing single-burst (LC class 1) from multi-burst (LC class 2) bolides, CNEOS light-curve records have inherent limitations. These records often capture only the brightest segments of a bolide's luminous path, often omitting earlier or later phases of ablation. Moreover, the absence of satellite sensor specifications (e.g., spectral response, dynamic range, field-of-view) and undisclosed platform positioning precludes robust correction for parallax or viewing-angle effects. Although the reported velocities and altitudes may carry inherent uncertainties (Devillepoix et al., 2019; Hajduková et al., 2024), these do not impact the present study. Nonetheless, the light-curve data near peak brightness remains invaluable for determining whether a bolide undergoes a single dominant burst or multiple discrete fragmentation episodes. Even with partial coverage and uncertain sensor calibration, the light-curve signal provides essential information regarding the timing and energy release pattern, allowing classification into LC class 1 or LC class 2. Furthermore, when combined with a velocity vector (if available), the data allow for reconstruction of the approximate altitude span of peak brightness and fragmentation (Silber and Sawal, 2025), an invaluable constraint when evaluating both the shock-formation details and the potential for multi-episode bursts.

For LC class 1 events, most of the bolide's energy is released abruptly as a single disruption, typically transitioning from a near-cylindrical shock to a quasi-spherical expansion. Consequently, this configuration often yields moderate period–energy slopes (**Figure 12d**) because the limited fragmentation constrains shock complexity. In contrast, LC class 2 events tend to yield steeper slopes as they exhibit multiple bursts along the bolide's path, with each releasing a significant fraction of the total energy. From a shock-physics perspective, these overlapping bursts effectively stack quasi-spherical expansions along the flight, broadening the low-frequency infrasound signature. Consequently, when bolide light-curve data are available, the relationships presented in Table 2 offer more reliable yield estimates.

### 4.4 Effects of Entry Geometry and Fragmentation on Period–Energy Relations

Bolide entry geometry strongly influences shock development and propagation, thereby shaping the $\log(E)$–$\log(P)$ relationship from our regression analyses. Fundamentally, a bolide traveling steeply at low altitudes concentrates its energy in a narrow near-cylindrical or quasi-spherical shock region, resulting in sharper period-energy gradients. In contrast, shallow or high-altitude trajectories introduce more complex shock interactions and extended propagation paths that weaken the energy–period correlation. This observation that steeper angles and lower altitudes correspond to higher slopes supports the premise that geometry governs the conversion of kinetic energy into infrasound. When a meteoroid enters at a steep angle ($\theta > 60°$) and penetrates to relatively low altitudes, it deposits substantial energy into a narrow vertical column (see **Figure 2**). The shock may initially resemble a cylindrical line-source and then transition into a quasi-spherical expansion with abrupt fragmentation, yielding a larger blast radius and stronger coupling between energy and period. Our regression fits under these conditions indicate that moderate energy increases correspond to notably longer acoustic periods. In contrast, shallow-angle entries ($\theta < 30°$) typically disperse energy over a longer horizontal path, with multiple trajectory segments generating overlapping shock fronts, especially when multi-fragmentation occurs. This elongated geometry leads to partial wave interference and more gradual 'smearing' of higher-frequency shock components, thus reducing the observed slope in period–energy space. Altitude further modulates these effects: deeper bursts (< 30 km altitude) frequently behave like localized explosions, whereas higher-altitude fragmentation often allows greater shock attenuation.

The distinction between bolides and nuclear explosions is apparent here. Assuming a single point-source explosion for bolides may be misleading in high-energy, shallow-entry cases such as the Chelyabinsk superbolide (Pilger et al., 2015; Popova et al., 2013; Silber, 2024a, 2025). Traditional infrasound energy–period relations based on point-source models do not capture the complex shock geometries resulting from multi-fragmentation or extended atmospheric paths. Recent work by Silber (2025) demonstrates that shallow angles can result in significant station-to-station variability in observed back azimuths (exceeding 10° over thousands of kilometers) because acoustic energy emerges from multiple points along the bolide trail rather than from a single point. This multi-segment emission can lead to substantial discrepancies between observed and modeled back azimuths when a unique peak brightness or point-source location is assumed (Silber, 2024a). Under favorable energetic and geometric conditions, anisotropy in acoustic emissions may occur, as observed in the Chelyabinsk event (Pilger et al., 2015).

## 4.5 Influence of Bolide Physical Parameters

Dense, cohesive bolides and those with high dynamic strength are negatively correlated with altitude (**Figure 12**), suggesting that they retain structural integrity until lower altitudes. In these conditions, energy is released into a denser atmospheric region, creating a more localized shock that intensifies wave amplitude and period. Velocity shows only moderate correlations with energy or size, suggesting that a wide range of orbital parameters can obscure simple mass or composition effects. Most meteoroid physical properties (strength, shape, density, and size) control the ablation and fragmentation rates, and therefore the energy deposition profile that generates the detectable infrasound signal. Furthermore, the expansion characteristics of the ablated vapor, particularly from the non-condensable volatile component, influence the shock radius and period. This is analogous to the Lamb wave coupling efficiency proposed by Boslough and Titov (2024), which is suggested to be greater for explosive volcanic events associated with volatile-rich magmas, and impact cratering events into volatile-rich target rocks. Overall, our results indicate that no single physical property dominantly governs the infrasound period. Nonlinear processes during hypervelocity entry (e.g., orientation-dependent drag, the response of internal structure to stress, and sequential fragmentation dynamics) contribute to the complex interplay that determines the shock wave characteristics.

### 4.6 Dependency on Range

One evident pattern from both the bootstrap method and OLS is that shorter ranges (generally ≤ 3000 km) often yield steeper slopes (*A*) compared to the global average (**Figure 11**), confirming that this effect is robust rather than resulting from a few outliers. In principle, more energetic bolides produce lower frequency infrasound components. As a result, these more powerful events can be detected at longer ranges (**Figure 13**) because low-frequency energy propagates efficiently through the atmosphere without substantial dissipative losses (Evans et al., 1972; Sutherland and Bass, 2004). This selectivity means that many far-range detections in large-area datasets often come from bolides with enough energetic output to maintain a discernible infrasound signal over thousands of kilometers. In far-field propagation, the initially complex shock can become effectively smoothed into its dominant low-frequency component. Consequently, once predominantly low-frequency content is all that remains, the measured periods do not vary as sharply with changes in bolide energy as they might at closer ranges. Thus, the observed period at long ranges can exhibit a slightly better correlation with energy than closer stations, where a fuller frequency spectrum is measured. While this ensures detectability for major bolides, it can also mask or average out certain mid- to high-frequency indicators of the event's energy, such as details linked to multiple bursts or rapid fragmentation. Our detection-range relation Eq. (6) diverges somewhat from that of Ens et al. (2012), likely because our larger, more diverse dataset includes additional low-energy bolides not captured in earlier work, resulting in longer average detection ranges for modest-energy events.

### 4.7 Are Period-Based Energy Relations Reliable?

A key outcome from our analysis is that no single global relation fully encapsulates the entire range of bolide infrasound behavior. While a single global line provides a convenient baseline, caution is warranted for nearby, localized events, where shock coupling can be stronger or more variable. Moreover, our findings show that while the classical slopes and intercepts from global fits remain practically useful, a parameter-aware approach is warranted when angles are shallow, altitudes are low, fragmentation is multi-episodic, or station ranges are especially short or long. Single-value relations are effective for initial energy estimates but may not capture extremes in geometric configuration or shock dynamics. Refining these relations through sub-group–specific fits (**Tables 2 and 3**) can improve the accuracy of infrasound-based energy assessment for both planetary defense applications and scientific investigations. Even if little else is known about a bolide beyond its infrasound record, our results show that period measurements alone can still yield a reasonable first approximation of total energy, and if multi-station records are available, geolocation. Whenever a station network detects sufficiently strong infrasound signals, regardless of uncertainties about flight path or fragmentation mode, one can employ established period–energy formulas to gauge yield. The effective propagation of low-frequency infrasound over large distances renders it a primary method for rapid bolide characterization, especially when optical or radar observations are lacking. Furthermore, refining energy estimates through data partitioning (e.g., by distance or inferred multi-fragmentation) advances risk assessment when supplementary information on the object's physical attributes are unavailable.

### 4.8 Broader Implications for Planetary Defense

Our results have direct implications for planetary defense, where accurate bolide energy estimates are crucial for assessing potential risks from near-Earth asteroids (NEAs). Robust correlations among diameter, density, and dynamic strength with total energy reveal that larger, more cohesive bolides tend to penetrate deeper, resulting in lower-altitude breakups that increase the potential for surface damage, and having distinct infrasound signatures. These observations support the use of period measurements as a reliable, rapidly obtained parameter for energy estimates, even though secondary factors (such as entry angle and altitude) modulate the shock's behavior. Clarifying how multi-fragmentation (LC class 2) intensifies the period–energy link through overlapping quasi-spherical expansions reinforces the need to account for fragmentation in yield predictions, particularly for low-altitude, multi-burst events. Recognizing that no single factor dominates the infrasound signal emphasizes the importance of multi-parameter models incorporating density, dynamic strength, angle, and altitude. The correlations derived here provide a framework for planetary defense agencies to assess potential hazards from incoming NEAs by linking fundamental shock physics to practical infrasound diagnostics.

Future work may benefit from advanced models, including Bayesian hierarchical frameworks and high-fidelity wave-propagation codes, to address complexities such as multi-episode fragmentation and shallow-angle entries. Nonetheless, our bootstrap and partition methodology forms a practical basis that bridges classical period–energy formulas with the complexity of real bolide observations, meeting the needs of both planetary defense and meteor physics research.

## 5. Conclusions

We have conducted a comprehensive investigation into empirical period–energy relations for bolides. We compiled consolidated tables of previously published period- and amplitude-based energy relations, correcting minor unit inconsistencies or typographical issues that have persisted in the literature. By incorporating an expanded bolide database that includes USG light curve and atmospheric Doppler wind profile data, and by applying both classical OLS fits and a bootstrap framework that partitions data by various parameters (e.g., altitude, source-to-station distance, entry angle, bolide mass, and light-curve classification), we have moved beyond the limitations of a single, global relation. This methodology provides a more comprehensive, physics-consistent means of refining yield estimates than relying on any single, universal formula.

Our refined period–energy fits demonstrate that while global or averaged fits are acceptable baselines for distant or minimally constrained events, significant variability appears in cases of low altitude, close stations, or shallow trajectories. Under these conditions, the measured infrasound period becomes much more sensitive to energy, explaining why a single universal relation may under- or over-estimate yields for specific geometries. By systematically isolating factors such as peak brightness altitude, fragmentation type, and station distance, our study reveals the origins of deviations from classical period–energy lines. This partitioned, bootstrap-driven approach, rarely applied so extensively in bolide research, offers a transparent, data-driven method for quantifying uncertainties and testing physical assumptions in shock formation

From a planetary defense hazard-assessment perspective, our framework allows for adaptable yield estimations tailored to bolide geometry and fragmentation type. Although an averaged global fit remains suitable for first-order estimates, subgroup partitions become crucial for shallow entries, low altitudes, or multiple fragmentation episodes where period–energy scaling significantly deviates from a single baseline. Our findings confirm that period measurements alone can provide a practical first approximation of bolide energy, even without additional observational constraints, and that multi-station records enable geolocation. The demonstrated effectiveness of infrasound detection in characterizing large, potentially hazardous bolides across vast distances emphasizes its immense value for planetary defense.

## Data Availability

CNEOS fireballs database is available at: https://cneos.jpl.nasa.gov/fireballs/. Light curve data are available on the CNEOS webpage: https://cneos.jpl.nasa.gov/fireballs/lc/. Data tables for all events examined in this study, along with the accompanying bootstrap outputs, can be accessed through Harvard Dataverse, doi: 10.7910/DVN/P6XZ67.

## Funding

This work, in part, was supported by the Laboratory Directed Research and Development (LDRD) program (project number 229346) at Sandia National Laboratories, a multimission laboratory managed and operated by National Technology and Engineering Solutions of Sandia, LLC., a wholly owned subsidiary of Honeywell International, Inc., for the U.S. Department of Energy's National Nuclear Security Administration under contract DE-NA0003525. JMT-R acknowledges financial support from the Spanish research project PID2021-128062NB-I00 funded by MCIN/AEI/10.13039/501100011033. EP-A acknowledges support from the LUMIO project funded by the Agenzia Spaziale Italiana (2024-6-HH.0). IO acknowledges the Universities Research Association-Sandia Graduate Student Fellowship Program which supported this research. SC acknowledges support provided by NSF Award EAR-1852339 and the NSF SAGE and GAGE facilities, operated by EarthScope Consortium. PJ is supported by NASA award 80NSSC22K1467 and a grant from the NASA Asteroid Threat Assessment Project.

## Acknowledgements

Sandia National Laboratories is a multi-mission laboratory managed and operated by National Technology and Engineering Solutions of Sandia, LLC (NTESS), a wholly owned subsidiary of Honeywell International Inc., for the U.S. Department of Energy's National Nuclear Security Administration (DOE/NNSA) under contract DE-NA0003525. This article has been authored by an employee of National Technology & Engineering Solutions of Sandia, LLC under Contract No. DE-NA0003525 with the U.S. Department of Energy (DOE). The employee owns all right, title and interest in and to the article and is solely responsible for its contents. The United States Government retains and the publisher, by accepting the article for publication, acknowledges that the United States Government retains a non-exclusive, paid-up, irrevocable, world-wide license to publish or reproduce the published form of this article or allow others to do so, for United States Government purposes. The DOE will provide public access to these results of federally sponsored research in accordance with the DOE Public Access Plan https://www.energy.gov/downloads/doe-public-access-plan. This paper describes objective technical results and analysis. Any subjective views or opinions that might be expressed in the paper do not necessarily represent the views of the U.S. Department of Energy or the United States Government. This article has been authored by an employee or employees of Triad National Security, LLC (Triad) under contract with the U.S. Department of Energy (DOE). Accordingly, the U.S. Government retains an irrevocable, nonexclusive, royalty-free license to publish, translate, reproduce, use, or dispose of the published form of the work and to authorize others to do the same for U.S. Government purposes.

## Declaration of competing interests

The authors declare no competing interests.

## Appendix A

### A.1. Empirical Amplitude-Range Yield Relations

In this section, we provide a table with empirical amplitude-range yield relations (Table A1). These relations are in the following form:

$$\log(E) = A \log(A^*) + B \log(R^*) + C,$$

where $E$ is energy in kilotons (or tons) of TNT equivalent, $P$ is infrasound signal period in seconds, and $A$, $B$ and $C$ are the regression coefficients, $A^*$ is the maximum amplitude ($A_{max}$) or wind-corrected maximum amplitude ($A_w$) in Pa, and $R^*$ is the range from the source to receiver in either km ($R$) or degrees ($\Delta$). The amplitude correction, derived by Whitaker (1995), is:

$$A_w = 10^{-0.019v} A_{max}.$$

**Table A1:** List of amplitude-range relations. Range ($R^*$) is listed in km or degrees, while $A^*$ refers to either maximum amplitude ($A_{max}$) or wind-corrected amplitude ($A_w$) in Pa. Stevens et al. (2002) included all arrivals, including thermospheric.

| A | B | C | $A^*$ [Pa] | $R^*$ | Source |
|---|---|---|---|---|---|
| 1.00 | 0.50 | -1.54 | $A_{max}$ | $R \sin \Delta$ | Pierce and Posey (1971) |
| 2.00 | 2.94 | -1.84 | $A_{max}$ | $\Delta$ | Clauter and Blandford (1998) |
| 1.47 | 2.00 | -4.96 | $A_{max}$ | $R$ | Whitaker (1995) |
| 3.03 | 3.03 | -9.09 | $A_{max}$ | $R$ | Russian-crosswind, Stevens et al. (2002) |
| 3.03 | 3.03 | -10.00 | $A_{max}$ | $R$ | Russian-downwind, Stevens et al. (2002) |
| 2.00 | 3.52 | -10.62 | $A_{max}$ | $R$ | Blanc et al. (1997) |
| 1.49 | 2.00 | -4.18 | $A_w$ | $R$ | Mutschlecner and Whitaker (2009) |
| 1.55 | 2.00 | -8.45 | $A_w$ | $R$ | Davidson and Whitaker (1992) |

### A.2 Amplitude-Range (Wind Corrected) Relations

There is a family of amplitude–range relations that incorporate an empirical correction constant ($k$) accounting for the average wind velocity ($v_W$) (in m/s) along the great-circle path from source to receiver (Tables A2 and A3). In these expressions, the signal amplitude ($A^*$) is given either as maximum amplitude ($A_{max}$) or peak-to-peak amplitude ($A_{P2P}$) in Pa:

$$\log(A^*) = a + b \log(R) + c \log(E) - k v_W.$$

Here, $a$, $b$, and $c$ are the regression coefficients, range ($R$) is in km, and the constant $k$ is expressed in units of (s/m). Half-yield amplitude-range relations are of the following form, where the coefficient $c$ is replaced by 0.5:

$$\log(A^*) = a + b \log(R) + 0.5 \log(E) - k v_W.$$

**Table A2:** List of amplitude-range relations that use wind correction. Range ($R$) is listed in km, while $A^*$ refers to either maximum amplitude ($A_{max}$) or peak-to-peak amplitude ($A_{P2P}$) in Pa. Ens et al. (2012) and Edwards (2007) derived their expressions from bolide observations. Mutschlecner et al. (1999) worked with nuclear test data, and Reed (1972) used high-yield explosive tests. Some relations are expressed in tons (t), whereas others are in kilotons (kt), as indicated in the "Units" column.

| Type | A* [Pa] | a | b | c | k [s/m] | Units | Source |
|------|---------|---|---|---|---------|-------|--------|
| <3.5 kt | $A_{max}$ | 4.06 | -0.99 | -0.43 | -0.0084 | t | Ens et al. (2012) |
| <3.5 kt | $A_{P2P}$ | 4.33 | -1.00 | -0.43 | -0.0084 | t | Ens et al. (2012) |
| >7 kt | $A_{max}$ | 10.75 | -1.15 | 0.78 | 0.0014 | t | Ens et al. (2012) |
| >7 kt | $A_{P2P}$ | 11.35 | -1.20 | 0.77 | 0.0013 | t | Ens et al. (2012) |
| all E, all R | $A_{max}$ | 4.38 | -1.06 | -0.47 | -0.0068 | t | Ens et al. (2012) |
| all E, all R | $A_{P2P}$ | 4.71 | -1.08 | -0.46 | -0.0068 | t | Ens et al. (2012) |
| <3000 km | $A_{max}$ | 4.97 | -1.18 | -0.45 | -0.0086 | t | Ens et al. (2012) |
| <3000 km | $A_{P2P}$ | 5.30 | -1.2 | -0.45 | -0.0084 | t | Ens et al. (2012) |
| <3.5 kt | $A_{max}$ | 3.21 | -1.75 | -- | -0.0174 | kt | Edwards (2007) |
| <3.5 kt | $A_{P2P}$ | 3.36 | -1.74 | -- | -0.0177 | kt | Edwards (2007) |
| >7 kt | $A_{max}$ | 2.18 | -1.26 | -- | -0.0024 | kt | Edwards (2007) |
| >7 kt | $A_{P2P}$ | 2.58 | -1.35 | -- | -0.0018 | kt | Edwards (2007) |
| all | $A_{max}$ | -- | -- | -0.46 | -- | kt | Mutschlecner et al. (1999) |
| all | $A_{max}$ | -- | -1.28 | -0.43 | -- | kt | Reed (1972) |

**Table A3:** List of half-yield amplitude-range relations that use wind correction. Range ($R$) is listed in km, while amplitude refers to either maximum amplitude ($A_{max}$) or peak-to-peak amplitude ($A_{P2P}$) in Pa. Some relations are expressed in tons (t), whereas others are in kilotons (kt), as indicated in the "Units" column.

| Type | Amplitude [Pa] | a | b | K [s/m] | Units | Source |
|------|----------------|---|---|---------|-------|--------|
| <3.5 kt | $A_{max}$ | 3.55 | -0.92 | -0.0083 | t | Ens et al. (2012) |
| <3.5 kt | $A_{P2P}$ | 3.79 | -0.93 | -0.0082 | t | Ens et al. (2012) |
| >7 kt | $A_{max}$ | 5.79 | -1.39 | -0.0006 | t | Ens et al. (2012) |
| >7 kt | $A_{P2P}$ | 6.24 | -1.44 | -0.0008 | t | Ens et al. (2012) |
| all E, all R | $A_{max}$ | 4.04 | -1.01 | -0.0068 | t | Ens et al. (2012) |
| all E, all R | $A_{P2P}$ | 4.28 | -1.02 | -0.0068 | t | Ens et al. (2012) |
| <3000 km | $A_{max}$ | 4.44 | -1.10 | -0.0086 | t | Ens et al. (2012) |
| <3000 km | $A_{P2P}$ | 4.68 | -1.11 | -0.0084 | t | Ens et al. (2012) |
| <3.5 kt | $A_{max}$ | 3.21 | -1.75 | -0.0174 | kt | Edwards (2007) |
| <3.5 kt | $A_{P2P}$ | 3.36 | -1.74 | -0.0177 | kt | Edwards (2007) |
| >7 kt | $A_{max}$ | 2.18 | -1.26 | -0.0024 | kt | Edwards (2007) |
| >7 kt | $A_{P2P}$ | 2.58 | -1.35 | -0.0018 | kt | Edwards (2007) |

## Appendix B

The infrasound signal periods used in this study were compiled from published data sources (Ens et al., 2012; Gi and Brown, 2017; Hupe et al., 2024; Ott et al., 2021; Ott et al., 2019; Pilger et al., 2020; Silber et al., 2024; Silber et al., 2011). The detailed procedures for measuring infrasound periods are described in Ens et al. (2012) and Silber (2024a). In essence, these methods identify the main segment of the waveform corresponding to the peak amplitude, then measure successive zero crossings to derive one or more fundamental periods. The dominant period is then computed from these zero-crossing intervals, often cross-validated by a spectral analysis to confirm the peak frequency (see Silber (2024a)). An illustrative example of this measurement process is shown in **Figure B1**, demonstrating how the maximum and minimum amplitudes are located, followed by consecutive zero crossings (ZC) that yield the signal period. Readers interested in a comprehensive discussion of the signal processing steps, potential sources of uncertainty, and station-to-station variability are referred to the above references. Because the measurements used in this work were taken directly from the literature, we do not replicate the entire procedure here.

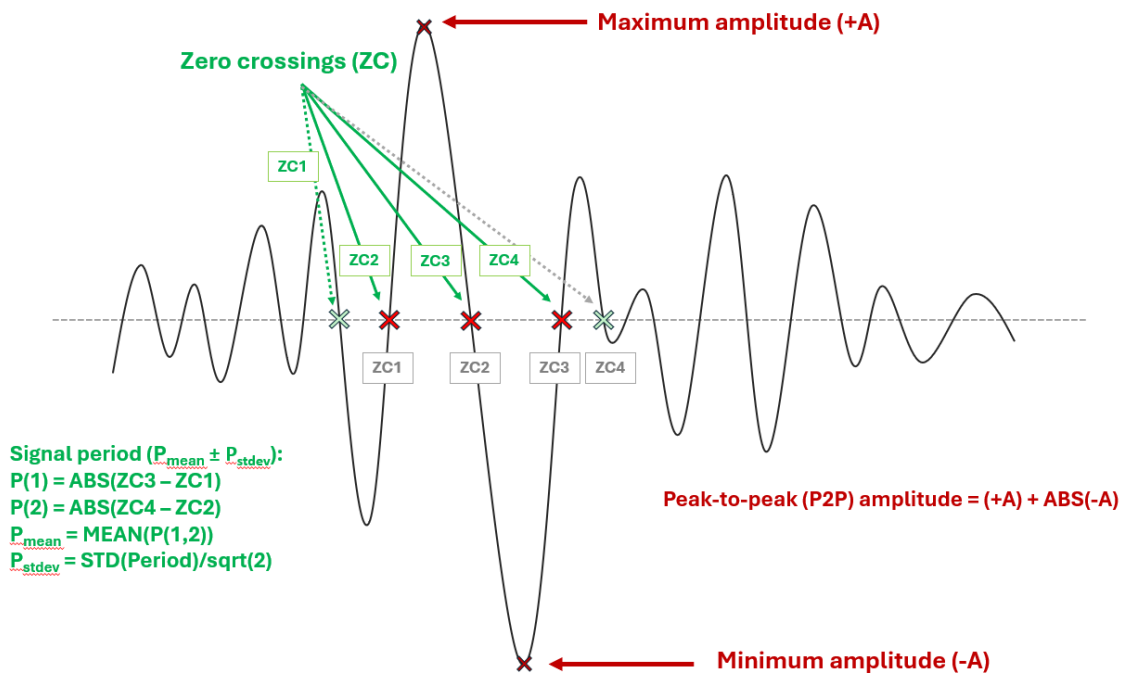

**Figure B1:** Schematic representation of the procedure for measuring infrasound signal parameters and quantifying the signal period. Each zero crossing (ZC) is marked where the waveform crosses the zero-pressure line. In this example, zero crossings labeled as ZC1, ZC2, ZC3, and ZC4 define the segments used to calculate the period. Notably, based on the analyst's judgment and the appearance of the waveform, ZC1 may be selected either before the maximum amplitude or at its onset. The green and gray sets of ZCs illustrate these possible scenarios. This flexibility in choosing ZC1 ensures that the extracted period accurately reflects the characteristics of the shock signal and accommodates variations in waveform morphology. Full-cycle ZCs define discrete periods. Averaging these discrete periods yields the mean signal period, and the standard deviation provides an uncertainty estimate. The maximum and minimum amplitudes (±A) are labeled, and the peak-to-peak (P2P) amplitude is computed as (+A)+|−A|. This method enables consistency when extracting signal parameters such as amplitude, period, and their statistical variability.